\documentclass{jfm}

\usepackage{graphicx}
\usepackage{newtxtext}
\usepackage{newtxmath}
\usepackage{natbib}
\usepackage{hyperref}
\hypersetup{
    colorlinks = true,
    urlcolor   = blue,
    citecolor  = black,
}

\newcommand{\RomanNumeralCaps}[1]

\newcommand{\rTSE}{r\overline{\mathrm{TSE}}}
\newcommand{\ralpha}{\mathrm{r\alpha}}

\title{Relative Fluid Stretching and Rotation for Sparse Trajectory Observations}

\author{Nikolas O. Aksamit\aff{1}
  \corresp{\email{nikolas.aksamit@uit.no}},
  Alex P. Encinas-Bartos\aff{2},
  George Haller\aff{2}
 \and David E. Rival\aff{3}}

\affiliation{\aff{1}Department of Mathematics and Statistics, UiT - The Arctic University of Norway, Forskningsparken 1, Tromsø 9037, Norway
\aff{2}Institute for Mechanical Systems, Swiss Federal Institute of Technology (ETH) - Zürich, Leonhardstrasse 21, Zürich 8092, Switzerland
\aff{3}Institute of Fluid Mechanics, Technische Universität Braunschweig, Hermann-Blenk-Strasse 37, Braunschweig 38108, Germany}

\begin{document}
\maketitle

\begin{abstract}
As most mathematically justifiable Lagrangian coherent structure detection methods rely on spatial derivatives, their applicability to sparse trajectory data has been limited. For experimental fluid dynamicists and natural scientists working with Lagrangian trajectory data via passive tracers in unsteady flows (e.g. Lagrangian particle tracking or ocean buoys), obtaining material measures of fluid rotation or stretching is currently only possible for trajectory concentrations that are often out-of-reach. To facilitate frame-indifferent investigations in unsteady and sparsely sampled flows, we present a novel approach to quantify fluid stretching and rotation via relative Lagrangian velocities. This technique provides a formal objective extension of quasi-objective metrics to unsteady flows by accounting for mean flow behavior. For extremely sparse experimental data, fluid structures may be significantly undersampled, and the mean flow behavior becomes difficult to quantify. We provide a means to maintain the accuracy of our novel sparse flow diagnostics in extremely sparse sampling scenarios, such as ocean buoy data and Lagrangian particle tracking. We use data from multiple numerical and experimental flows to show that our methods can identify structures beyond existing limits of sparse, frame-indifferent diagnostics, and exhibit improved interpretability over common frame-dependent diagnostics.
\end{abstract}

\section{\label{sec:Intro}Introduction}
Experimental methods for spatially and temporally resolving fluid velocities in both large and natural flow domains have improved significantly in recent years. For a wide range of turbulent flows, ground truth measurements of fluid velocity can be measured using high-speed imaging of advected particles and any number of particle image velocimetry (PIV) and Lagrangian particle tracking (LPT) algorithms. For the largest spatial scales and for the sparsest trajectory data, only LPT is suitable, and provides a Lagrangian framework with which to extract transport features in the flow. At oceanographic and atmospheric circulation spatial scales, GPS-tracking of buoys and balloons typically replaces PIV and LPT imaging approaches. To date, Lagrangian data has provided great insights to our studies of sea ice and ocean dynamics, as well as for meteorologists studying atmospheric behaviors with drifting weather-balloon measurements \citep{Businger2006, Lepparanta2011, VanSebille2018}. Coherent structure identification from sparse data, however, has typically relied on a grab-bag of techniques, often tailored to each individual flow, with no unifying metrics that work in all domains.

Historically, lab-based measurement techniques have evolved hand-in-hand with systematic technological advances such as integrated circuits, the laser, and most recently the CMOS chip. Starting out from intrusive, probe-based extraction of Eulerian data (fixed-point statistics) through to time-averaged planar field measurements, e.g. PIV, and then most recently to dense time-resolved particle tracking in three dimensions, the availability of said tools has influenced the choice of metrics used to describe the flow in question. As a case in point, the use of Reynolds stresses to describe shear flows has dominated the community since the days of hot-wire anemometry even though such stresses are only a proxy to the coherent structures driving the turbulent processes on hand.

Commonly used approaches for identifying structures in experimental flows are typically frame-dependent. That is, the extracted features will depend on the choice of reference frame of the experimentalist, and thus violate a fundamental requirement from continuum mechanics for describing \textit{material} fluid behavior.  Material behavior can be thought of as the features in a flow revealed in a tracer visualization experiment (e.g. dye, smoke, etc.). While our physical intuition around Lagrangian velocities may be strong, e.g., one can easily imagine a leaf floating downstream on a river surface, extracting physically meaningful diagnostics that describe the material deformation of the surrounding fluid is much more difficult. Indeed many common and intuitive trajectory metrics are frame-dependent, such as the Lagrangian velocity, looping \citep{Lumpkin2016}, curvature \citep{Bristow2023}, complexity measures \citep{Rypina2011}, and network-based approaches \citep{Iacobello2023}, as well as many diagnostics from gridded velocity data such as vorticity \citep{Bernard1993}, swirling strength \citep{Zhou1999}, and the lambda-2 criterion \citep{Jeong1995} . 

While choosing a single common reference frame may appear logical to study flows we understand well, this approach quickly loses its foundation when encountering flows where no \emph{a priori} knowledge of the relevant structures or a natural reference frame are available. Furthermore, truly material behavior does not change under rigid-body frame changes, regardless of whether a user-preferred frame exists or not \citep{Truesdell2004}. Thus Euclidean frame-indifference is a fundamental litmus test for structure identification schemes to actually identify material features even if one always conducts their research in the same reference frame.

Methods that can effectively reveal fluid structures at both high and low trajectory densities, in a frame-independent manner, and for the widest variety of flows would provide a wealth of information for both understanding fluids in observational environments and as a common ground for comparison with numerical simulations. In order to test sparse diagnostics in very different flows, we will quantify the sparsity of a dataset by normalizing the number of trajectories by the square or cube of a characteristic length scale of the flow $\ell$, for 2D and 3D examples, respectively. 

Frame-indifferent (objective) diagnostics that identify Lagrangian coherent structures have been extensively developed over the last two decades \citep[see][]{Haller2023} but their implementation relies on spatial derivatives that are difficult to accurately compute from sparse or unstructured data. To account for this, adaptations have been developed to allow for data sparsity \citep{Lekien2010, Rypina2021, Mowlavi2022} as have Green's-theorem-based approximations of rate-of-strain metrics from trajectory arrays \citep{Kwok1990}. Most notably, \citet{Mowlavi2022} compared multiple sparse methods for identifying hyperbolic (stretching) and elliptic (rotating) Lagrangian coherent structures in the Bickley jet and ABC flow. When initializing particles on a structured grid, \citet{Mowlavi2022} were able to accurately cluster particles into elliptic LCS at particle concentrations of 85$\ell^{-2}$ and 504$\ell^{-3}$ for the Bickley jet and ABC flow ($\ell=\pi$ for both), respectively. They were also able to identify hyperbolic structures from randomly initialized particles at concentrations of 471$\ell^{-2}$, and 3870$\ell^{-3}$ for the Bickley jet and ABC flow, respectively.

Modern clustering methods provide a complementary frame-indifferent approach to identify regions of fluid with similar fluid particle trajectories \citep[see, e.g.,][]{Froyland2015, Hadjighasem2016, Schlueter-Kuck2017}. Graph-theory-based clustering algorithms quantify similarity between trajectories themselves, and thus do not require measurements to be spatially proximal, as is required for gradient-reliant approaches. As such, trajectories can be generated from gridded flow data or be experimentally observed, such as by LPT. In contrast to hyperbolic and elliptic LCS, the coherent structures identified with trajectory clustering algorithms have no inherent physical meaning besides a certain trajectory similarity. That is, one cannot use clustering alone to interpret local stretching or rotation rates without {\it a priori} knowledge of the flow behavior.

The squared relative dispersion ($d^2$) is an outlier diagnostic as it is suitable for sparse and randomly oriented trajectories, is also objective \citep{Haller2000}, and has been developed with a strong physical foundation. This fluid stretching metric has been used for a number of years, particularly to understand dispersion and mixing by oceanographers in a statistical manner \citep{LaCasce2008}. The ability of $d^2$ to identify coherent structures, however, is limited as one is forced to initially choose particle pairs, whose relative motion inevitably becomes uncorrelated at an \emph{a priori} unknown temporal horizon \citep{Haller2021}.

Current approaches for physically meaningful sparse trajectory diagnostics still rely on a relatively dense field of particles \citep[for a comparison, see, e.g.,][]{Mowlavi2022}, or well-behaved trajectory array geometries \citep{Lindsay2003}.  In light of the unstructured nature of trajectory data, it is more common for experimental LPT data to be spatially and temporally averaged or assimilated into gridded products prior to structure analysis \citep{Schroder2023}. When averaging or projecting time-dependent fluid motions to a single instance in time, such approaches potentially oversimplify transient (nonstationary) flow behavior that may exist in the underlying Lagrangian data.

Complementing these sparse approaches is the recent development of quasi-objective coherent structure diagnostics, the trajectory rotation average (TRA) and trajectory stretching exponent (TSE) \citep{Haller2021}. TSE and TRA calculate stretching and rotation, respectively, for individual particle trajectories with no requirement of nearby velocity or trajectory data. The authors mathematically proved that under suitable conditions (slowly varying, relatively small mean vorticity) TRA and TSE approximate objective measures of rotation and stretching. Given these flow conditions, TRA and TSE have proven to be advantageous in several extremely sparse geophysical buoy experiments, accurately identifying algae trapping eddies in the ocean \citep{Encinas-Bartos2022}, as well as predicting Arctic sea ice stretching and breakup events that were missed by other approaches \citep{Aksamit2023a}.

In the present research, we introduce relative stretching and rotation metrics for individual trajectories that incorporate knowledge of the average translation and rotation of the sampled fluid. By utilizing bulk behavior of concurrent trajectories in a given experiment, we can obtain objective diagnostics of stretching and rotation in unsteady flows with much of the same flexibility of a true single-trajectory method. Our measures of relative stretching and rotation can be seen as a natural synthesis of the objective Eulerian deformation velocity of \citet{Kaszas2023} and quasi-objective diagnostics of \citet{Haller2021}. 

In the following, we develop the theory of relative stretching and rotation and provide several examples of performance. The robustness of relative Lagrangian stretching and rotation is displayed by first comparing full resolution structure topology against ground truth LCS boundaries in 2D and 3D flows, and then testing the accuracy of the trajectory diagnostics in progressively downsampled datasets. We also display the enhanced ability of relative rotation to distinguish between experimental turbulent flows for extremely sparsely sampled data, when compared with traditional metrics. This superior performance is shown for numerical simulation data and in real world observations of ocean buoys from the global drifter database and a large-scale LPT wind tunnel experiment.

\section{Methods}
\subsection{Background}
Consider a fluid flow with time-varying velocity field $\mathbf{v}(\mathbf{x},t)$. Infinitesimal fluid particle trajectories can be generated as solutions to the differential equation $\dot{\boldsymbol{x}}(t)=\mathbf{v}(\boldsymbol{x}(t),t)$ from some initial position $\boldsymbol{x}(t_0)=\boldsymbol{x}_0$. The flow map $\mathbf{F}$ maps fluid particles from their time $t_0$-positions to their position at a time $t$, along the trajectory $\boldsymbol{x}(t; t_0, \boldsymbol{x}_0)$,

$$\mathbf{F}_{t_0}^t: \boldsymbol{x}_0\mapsto \boldsymbol{x}(t; t_0, \boldsymbol{x}_0).$$

The measurement of a scalar quantity associated to a fluid particle $P(\boldsymbol{x}(t))$, such as its temperature, is objective (frame-indifferent) under Euclidean transformations of the form 
\begin{equation}
\mathbf{x}=\mathbf{Q}(t)\mathbf{y}+\mathbf{b}(t),
\label{eq: Transformation}
\end{equation}
where $\mathbf{Q}(t)$ is a rigid body rotation matrix (an element of the matrix group $\mathrm{SO}(n)$ for $n=2$ or $3$) and $\mathbf{b}(t)$ is a time-varying translation vector. That is, in the two reference frames, the scalar quantity remains the same: $\tilde{P}(\boldsymbol{y})=P(\boldsymbol{x})$, where $P$ and $\tilde{P}$ are measured in the original and translated frames, respectively. 

Similarly, one can define an objective Lagrangian vector $\boldsymbol{\xi}$ as one that transforms under eq (\ref{eq: Transformation}) as $$\tilde{\boldsymbol{\xi}}(\boldsymbol{y}(t,\boldsymbol{y}_0)) = \mathbf{Q}^T(t)\boldsymbol{\xi}(\boldsymbol{x}(t,\boldsymbol{x}_0)).$$ 

To reveal the ground truth material fluid structures undergoing significant stretching and rotation, we will rely on the frame-indifferent finite-time Lyapunov exponent (FTLE) \citep{Haller2015} and the Lagrangian-averaged vorticity deviation (LAVD) \citep{Haller2016}, respectively:

\begin{align}
\mathrm{FTLE}_{t_{0}}^{t}(\mathbf{x}_{0})&=\frac{1}{2\left|t-t_{0}\right|}\log\lambda_{\mathrm{max}}\left(\mathbf{C}_{t_0}^t\left(\mathbf{x}_{0}\right)\right),\label{eq:FTLE field}\\
\mathrm{LAVD}_{t_{0}}^{t}(\mathbf{x}_{0})&=\frac{1}{|t-t_{0}|}\int_{t_{0}}^{t}\left|\boldsymbol{\omega}(\mathbf{F}_{t_0}^s(\mathbf{x}_{0}),s)-\bar{\boldsymbol{\omega}}(s)\right|ds,
\end{align}
where $\lambda_{\mathrm{max}}>0$ is the largest eigenvalue of the positive definite tensor $\mathbf{C}_{t_0}^t=\left[\boldsymbol{\nabla}\mathbf{F}_{t_0}^t\right]^{\mathrm{T}}\boldsymbol{\nabla}\mathbf{F}_{t_0}^t$ and $\bar{\boldsymbol{\omega}}$ is the time-varying but spatially averaged vorticity. While LAVD and FLTE have seen widespread use in studies of atmospheric, oceanic, and experimental flows, their direct application to sparse datasets has been hindered by their strong reliance on velocity and flow map gradients.

In the sparse sampling context, where we have no control over trajectory densities or orientations, it would be advantageous to develop an LCS algorithm based on individual trajectories. The primary issue surrounding the development of an objective single-trajectory diagnostic is that one can always pass to the frame of the particle, and in that reference frame, the particle is not moving. As a way around this frame-dependency, \citet{Haller2021} recently developed the quasi-objective coherent structure diagnostics 

\begin{align}
\mathrm{\overline{TSE}}_{t_0}^{t_N}(\boldsymbol{x}_0)&=\frac{1}{t_N-t_0}\sum_{j=0}^{N-1}\left |\log\frac{|{\dot{\boldsymbol{x}}}(t_{j+1})|}{|{\dot{\boldsymbol{x}}}(t_{j})|} \right|, \label{Eq: TSE}\\
 \overline{\mathrm{TRA}}_{t_0}^{t_N}(\boldsymbol{x}_0) &= \frac{1}{t_N-t_0}\sum_{j=0}^{N-1}\cos^{-1}\frac{\langle \dot{\boldsymbol{x}}(t_j), \dot{\boldsymbol{x}}(t_{j+1}) \rangle} {|\dot{\boldsymbol{x}}(t_j)| |\dot{\boldsymbol{x}}(t_{j+1})|},
\end{align}
which provide close approximations to objective stretching and rotation measures \emph{if} the Lagrangian trajectories are analyzed in slowly varying references frames with relatively small mean vorticity.

In the following, we derive the experiment-relative Lagrangian velocity of a fluid particle and show that it is an objective vector. With these relative velocities, we can objectively define the relative stretching and rotation of fluid from sparsely sampled experimental data, with no {\it a priori} knowledge of the structures being investigated.

\subsection{Relative Lagrangian Velocity}

Suppose we have a collection of Lagrangian particle trajectory observations $\{\boldsymbol{x}_i(t)\}$, with corresponding velocities along their trajectories $\dot{\boldsymbol{x}}_i(t)=\boldsymbol{v}_i(t)$. By linearity of the time-derivative, the velocity of the average position of these particles at time $t$ is equivalent to the average of their Lagrangian velocities,

\begin{equation}
\overline{\boldsymbol{x}}(t)=\frac{1}{n}\sum_{i=1}^n \boldsymbol{x}_i(t)
\label{Eq: Position Avg}
\end{equation}

\begin{equation}
\boldsymbol{v}(t)=\dot{\boldsymbol{x}}(t), \qquad  \overline{\boldsymbol{v}}(t)=\frac{1}{n}\sum_{i=1}^n \boldsymbol{v}_i(t)=\dot{\overline{\boldsymbol{x}}}(t)
\label{Eq: Velocity Avg}
\end{equation}

Furthermore, define the time-varying moment-of-inertia tensor $$\Theta(t)=\overline{|\boldsymbol{x}_i-\overline{\boldsymbol{x}}|^2I-(\boldsymbol{x}_i-\overline{\boldsymbol{x}})\otimes(\boldsymbol{x}_i-\overline{\boldsymbol{x}})}$$ where averaging is over the trajectory index $i$. From our collection of trajectories, mimicking the Eulerian calculations of  \citet{Kaszas2023}, we relate the experiment-based vorticity to the approximate rigid-body rotation of the flow as 
\begin{equation}
\overline{\boldsymbol{\omega}}_{rb}(t)=\Theta^{-1}\overline{(\boldsymbol{x}_i-\overline{\boldsymbol{x}})\times(\boldsymbol{v}_i-\overline{\boldsymbol{v}})}.
\label{Eq: Vort Avg}
\end{equation}

To avoid cumbersome notation, we will drop the subscript $i$ from hereon when it is clear we are referring to a specific trajectory $\boldsymbol{x}(t; t_0, \boldsymbol{x}_0)$. Having now calculated the average rotation and translation of the sampled fluid in our experiment, we define the relative velocity along a Lagrangian trajectory as 

\begin{equation}
\boldsymbol{v}_d(\boldsymbol{x}(t))=\boldsymbol{v}(\boldsymbol{x}(t))-\overline{\boldsymbol{v}}(t) - \overline{\boldsymbol{\omega}}_{rb}(t)\times(\boldsymbol{x}-\overline{\boldsymbol{x}}).
\label{eq: Relative LagVel}
\end{equation}
The objectivity of $\boldsymbol{v}_d$, and scalar diagnostics generated from it, is proven in Appendix A. That is, by removing the mean rate of translation and rotation of our trajectory observations, we can study objective properties of dynamic fluid structures in our sampled domain.

This is a Lagrangian-observation-based implementation of the Eulerian deformation velocity $\mathbf{v}_d$ originally derived by \citet{Kaszas2023}. This is, however, distinct from calculating Lagrangian trajectories in time-resolved Eulerian $\mathbf{v}_d$ fields. Trajectories calculated in the Eulerian $\mathbf{v}_d$ do not necessarily trace material flow features, but $\mathbf{v}_d$ can be used to objectivize frame-dependent Eulerian diagnostics in each snapshot. Here, by estimating rigid-body translation and rotation from passive Lagrangian measurements, the relative Lagrangian velocity $\boldsymbol{v}_d$ provides a means to objectivize trajectory diagnostics relative to the flow data available without {\it a priori} knowledge of the underlying flow. 

\subsection{Stretching}

For steady flows, \citet{Haller2021} showed that cumulative fluid stretching and compression normal to trajectories can be quantified using $\mathrm{\overline{TSE}}$ (\ref{Eq: TSE}). In slowly varying flows, this trajectory stretching and compression approximates the true material deformation with the difference being a function of $\partial_t\mathbf{v}(\boldsymbol{x}(t),t)$. Furthermore, TSEs were shown to faithfully highlight hyperbolic LCSs and reproduce the dominant features of FTLE fields.

Here, we seek a measure of fluid stretching that is frame-indifferent in unsteady flows and has minimal dependence on trajectory concentration, thus making it suitable for extremely sparse sampling in many natural environments. To do this, we introduce the stretching of fluid parcels relative to the bulk behavior of our flow through the stretching of relative Lagrangian velocity vectors. Formally, the {\it relative} trajectory stretching exponents rTSE and $\mathrm{r\overline{TSE}}$ from time $t_0$ to time $t_N$ can be written as

\begin{align}
\mathrm{rTSE}_{t_0}^{t_N}(\boldsymbol{x}_i) &=\frac{1}{t_N-t_0}\log\frac{|\boldsymbol{v}_d({\boldsymbol{x}}_i(t_N))|}{|\boldsymbol{v}_d({\boldsymbol{x}_i}(t_0))|} \label{rTSE},\\
\mathrm{r\overline{TSE}}_{t_0}^{t_N}(\boldsymbol{x}_i)&=\frac{1}{t_N-t_0}\sum_{j=0}^{N-1}\left |\log\frac{|\boldsymbol{v}_d({\boldsymbol{x}}_i(t_{j+1}))|}{|\boldsymbol{v}_d({\boldsymbol{x}}_i(t_{j}))|} \right|.
\label{rTSE_Bar}
\end{align}
The $\mathrm{rTSE}$ diagnostic provides the stretching exponent for the relative Lagrangian velocity vector from time $t_0$ to $t_N$, whereas $\mathrm{r\overline{TSE}}$ is a cumulative measure of all stretching and contraction that occurs in the same time window. Given the objectivity of the relative Lagrangian velocity $\boldsymbol{v}_d$ defined in eq (\ref{eq: Relative LagVel}), then $\mathrm{rTSE}$ and $\mathrm{r\overline{TSE}}$ are also objective. That is, because $\hat{\boldsymbol{v}}_d(\boldsymbol{y}(t))=Q^T\boldsymbol{v}_d(\boldsymbol{x}(t))$, we have $|\hat{\boldsymbol{v}}_d(\boldsymbol{y}(t))|=|\boldsymbol{v}_d(\boldsymbol{x}(t))|$ and rTSEs do not change between reference frames. This provides a formal and fully objective extension of the quasi-objective TSEs from \citet{Haller2021} that works in unsteady (time-varying) flows.

\subsection{Rotation}
\label{Sec: Rotation}
For steady flows with negligible mean vorticity, \citet{Haller2021} also derived the single trajectory rotation average ($ \overline{\mathrm{TRA}} $), which measures trajectory rotation and successfully identifies elliptic Lagrangian coherent structures. Under these flow assumptions, $ \overline{\mathrm{TRA}} $ calculates the cumulative rotation of material streamline tangent vectors. For highly unsteady, or strongly rotational flows this approach is insufficient as streamlines no longer resemble material lines. Instead we adopt the average rotation speed of material tangent vectors for relative Lagrangian velocities.

For a given relative Lagrangian velocity vector $\boldsymbol{v}_d$, we can define the unit vector pointing in that direction:

\begin{equation}
\mathbf{e}(t)=\frac{\boldsymbol{v}_d(t)}{| \boldsymbol{v}_d(t)|}. \label{Eq: Eprime}
\end{equation}
Then, 
\begin{equation}
\mathrm{r\alpha}_{t_0}^{t_N}=\frac{1}{t_N-t_0}\int_{t_0}^{t_N}\left| \dot{\mathbf{e}}(t) - \frac{1}{2}\bar{\boldsymbol{\omega}}_{rb}(t)\times\mathbf{e}(t)\right|dt
\label{Eq: Alpha}
\end{equation}
is an objective measure of the average rotation speed of $\boldsymbol{v}_d$ from time $t_0$ to $t_N$ \citep{Haller2021}. The instantaneous limit of eq (\ref{Eq: Alpha}) also exists ($\alpha_{t_0}=\lim_{t_N\to t_0}\alpha_{t_0}^{t_N}$) and provides the instantaneous rate of rotation of the relative Lagrangian velocity vector with respect to the spatially-averaged rotation. Furthermore, for a generic parameterized curve $\gamma$ in 3D with a unit tangent velocity vector $\mathbf{e}$, the curvature of $\gamma$ is $\dot{\mathbf{e}}(t)$. Thus $\ralpha$ is intrinsically related to classical notions of geometry of flow structures as a form of integrated curvature that accounts for rotation of the surrounding fluid as well.

\subsection{Extremely Sparse Sampling}
\label{Sec: Extreme}

As we will see in the following sections, calculating $\overline{\boldsymbol{v}}(t)$ and $\overline{\boldsymbol{\omega}}(t)$ directly from trajectory data can become problematic at extremely sparse trajectory concentrations. By extremely sparse, we refer to trajectory concentrations less than $1\ell^{-2}$ and $1\ell^{-3}$, where $\ell$ is a characteristic length of structures in the flow (e.g., eddy length scale, obstruction dimensions). At this lower limit of trajectory concentration, we are fundamentally undersampling the prominent structures in the flow, and an accurate time-resolved depiction of spatially-averaged flow behavior is not reasonably expected.

That is, $ \boldsymbol{\overline{\omega}}_{rb}(t) $ and $ \boldsymbol{\overline{v}}(t) $ may start to display strong oscillations in time that are not representative of the underlying mean flow properties. These oscillations contribute unphysical fluctuations in $ \boldsymbol{v}_{d}$ and hinder our interpretation of relative stretching and rotation diagnostics. An example of this is shown in detail in Section \ref{sec:AVISO}.

For these situations, we suggest replacing the time-dependent bulk values $ \boldsymbol{\overline{\omega}}_{rb},\boldsymbol{\overline{v}}, \boldsymbol{\overline{x}} $ from eqs (\ref{Eq: Position Avg}-\ref{Eq: Vort Avg}) with their temporal averages over the integration window as follows:
\begin{align}
\boldsymbol{v}_{\mathrm{ref}} & = \dfrac{1}{t_1-t_0}\int_{t_0}^{t_1} \boldsymbol{\overline{v}}(t)dt, \label{Eq: Time Avg v}\\ 
\boldsymbol{\omega}_{\mathrm{ref}} &= \dfrac{1}{t_1-t_0}\int_{t_0}^{t_1} \boldsymbol{\overline{\omega}}_{rb}(t)dt, \label{Eq: Time Avg w}\\
\boldsymbol{x}_{\mathrm{ref}} &= \dfrac{1}{t_1-t_0}\int_{t_0}^{t_1}\boldsymbol{\overline{x}}(t)dt. \label{Eq: Time Avg x}
\end{align} 

We can then write the Lagrangian velocity relative to this prescribed reference frame as
\begin{equation}
\boldsymbol{v}_{d\mathrm{,ref}}(\boldsymbol{x}(t))=\boldsymbol{v}(\boldsymbol{x}(t))-\boldsymbol{v}_{\mathrm{ref}}(t) - \boldsymbol{\omega}_{\mathrm{ref}}(t)\times(\boldsymbol{x}-\boldsymbol{x}_{\mathrm{ref}}),
\label{eq: Relative LagVel w/ Reference}
\end{equation}
where $\boldsymbol{v}_{\mathrm{ref}}$ and $\boldsymbol{\omega}_{\mathrm{ref}}$ are steady approximations of the bulk translation and rotation in the flow that are informed directly from flow measurements. This extremely sparse data handling is tested for multiple flows in the following sections, as is its ability to recreate accurate full-resolution $\ralpha$ values. Since we are explicitly keeping track of a given choice of reference frame in eq (\ref{eq: Relative LagVel w/ Reference}), $\boldsymbol{v}_{d\mathrm{,ref}}$, $\rTSE$ and $\ralpha$ remain objective and indifferent to Euclidean frame changes. 

If additional sources of velocity measurements are available, such as those from anemometers, remotely sensed ocean currents, or wind tunnel probes, one may also estimate $\boldsymbol{v}_{\mathrm{ref}}$ and $\boldsymbol{\omega}_{\mathrm{ref}}$ from those data, but this approach is not tested herein and should only be considered with caution. Using additional data sources to estimate bulk rigid-body rotation and translation is subject to errors, such as local fluctuations or imperfect placement in the flow, that are not discussed in the present work. 

If one imposes $\boldsymbol{v}_{\mathrm{ref}}=0$, and $\boldsymbol{\omega}_{\mathrm{ref}}=0$, we can see that TRA and TSE are actually special cases of $\mathrm{r}\alpha$ and $\mathrm{rTSE}$. The present work, therefore, seeks to extend the prior success found by \citet{Haller2021} and \citet{Encinas-Bartos2022} with TRA and TSE to unsteady, time-varying flows. Further, the generalized relative Lagrangian velocity frameworks allows researchers to transparently maintain objectivity if a given reference frame is preferred.
 
 \section{Results}
We now provide examples of $\mathrm{r\alpha}$ and $\mathrm{r\overline{TSE}}$ analysis for numerical simulations and experimental observations. We evaluate our newly proposed metrics against an array of diagnostics suitable for different experimental flows, with each example detailing a specific advantage of $\mathrm{r\alpha}$ and/or $\mathrm{r\overline{TSE}}$ over prior approaches.

\subsection{Numerical Simulations}

\subsubsection{Unsteady Bickley Jet}
\label{Section: Bickley}
For our first example we consider the unsteady Bickley jet. This is a two-dimensional geophysical model of a quasi-periodic zonal jet with adjacent migrating eddies whose Lagrangian dynamics have been studied in depth \citep[see, e.g.,][]{Del-Castillo-Negrete1993, Rypina2007}. This flow was also used as a benchmark for previous Lagrangian coherent structure comparisons \citep{Hadjighasem2017}. Its time-dependent stream function is given by

\begin{equation}
\psi(x,y,t) = -UL\mathrm{tanh}\left(\frac{y}{L}\right) + UL\mathrm{sech}^2\left(\frac{y}{L}\right)Re\left[\sum_{n=1}^3 \epsilon_n\exp(-ik_nc_nt)\exp(ik_nx)\right].
\end{equation}
We use the parameters from \citet{Rypina2007} found in Table \ref{Table: Bickley}.

\begin{table}
\begin{center}
\begin{tabular}{||c c c c c c c c c||} 
 \hline
 U & L & $k_n$ & $\epsilon_1$ & $\epsilon_2$ & $\epsilon_3$ & $c_3$ & $c_2$ & $c_1$ \\ [0.5ex] 
 \hline\hline
 62.66 $m$ $s^{-1}$ & 1770 $km$ & $2n/r_0$ & 0.0075 & 0.15 & 0.3 & $0.205U$ & $0.461U$ & $c_3+\frac{k_2(\sqrt{5}-1)}{2k_1}(c_2-c_3)$\\
 \hline
\end{tabular}
\caption{Bickley Jet model parameters.}
 \label{Table: Bickley}
\end{center}
\end{table}

\begin{figure}
\includegraphics[width=\textwidth]{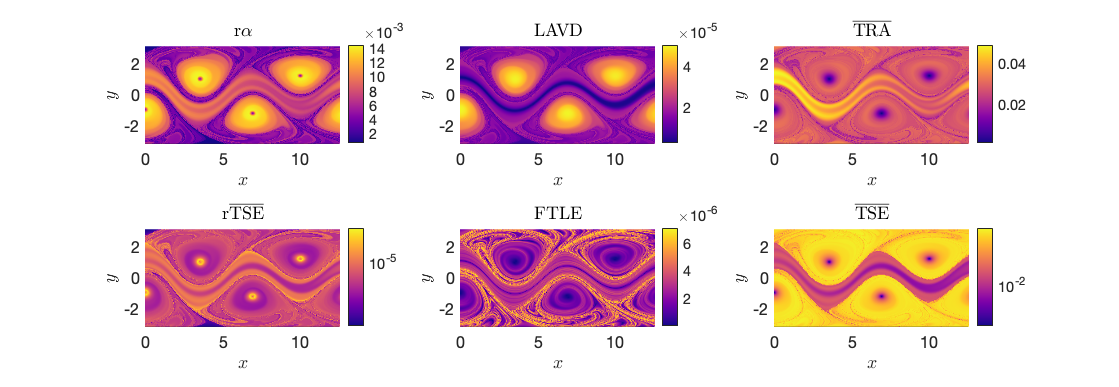}
\caption{Evaluation of $\mathrm{r}\alpha$ and $\mathrm{r\overline{TSE}}$ (left) with material rotation (LAVD) and stretching (FTLE) (center), and traditional quasi-objective single trajectory diagnostics $\overline{\mathrm{TRA}}$ and $\overline{\mathrm{TSE}}$, in the unsteady Bickley Jet. Note the distinct central jet with rotating eddies on either side. $\overline{\mathrm{TRA}}$ and $\overline{\mathrm{TSE}}$ have difficulty accurately capturing material behavior (stretching and rotation) due to the large bulk advection in the flow and temporal variability in the velocity field.}
\label{Fig: Bickley Stretch}
\end{figure}

In figure \ref{Fig: Bickley Stretch} we compare $\mathrm{r\alpha}$ and $\rTSE$ to benchmarks of material rotation and stretching, the Lagrangian averaged vorticity deviation (LAVD) \citep{Haller2016}, and the finite time Lyapnuov exponent (FTLE) \citep{Haller2015} on a dense grid of trajectories (3600$\ell^{-2}$, $\ell=\pi$), for an integration time of 30 days. This corresponds with approximately 6 full rotations and significant translation for the advected eddies. We find that $\mathrm{r\alpha}$ is able to match the prominent features of the LAVD field by effectively identifying eddies boundaries as closed convex contours and separating them from the central jet. $\rTSE$ ridges identify the edges of the central jet and edges of vortices as regions of significant stretching, similar to FTLE. Deviations primarily exist in the centers of the vortex cores where $\boldsymbol{v}_d(\boldsymbol{x}(t))$ does not evolve in a material way.

Previous work has shown how $\overline{\mathrm{TRA}}$ and $\overline{\mathrm{TSE}}$ can effectively reproduce the same LCS as LAVD and FTLE in slowly varying flows \citep{Haller2021}. In figure \ref{Fig: Bickley Stretch}, we reveal how necessary this slowly varying assumption truly is. With the parameters in Table \ref{Table: Bickley}, the Bickley jet is highly unsteady. For meaningful quasi-objective calculations, one requires $|\mathbf{v}_t|/|\ddot{\boldsymbol{x}}|<<1$, but only $41\%$ of trajectories experience $|\mathbf{v}_t|/|\ddot{\boldsymbol{x}}|<1$ and only $0.03\%$ of particle paths show $|\mathbf{v}_t|/|\ddot{\boldsymbol{x}}|<0.1$. This suggests that the coherent structures in this flow are likely evolving or traveling much faster than a Lagrangian particle is able to trace them out. In fact, $\overline{\mathrm{TRA}}$ designates the eddies in figure \ref{Fig: Bickley Stretch} as exhibiting relatively low rotation. $\overline{\mathrm{TSE}}$ also suggests nearly everything outside the central jet is undergoing significant stretching, in contrast to the thin hyperbolic regions seen as FTLE ridges. For this degree of unsteadiness and strong eddy advection, $\mathrm{r\alpha}$ and $\rTSE$ are more suitable than $\overline{\mathrm{TRA}}$ and $\overline{\mathrm{TSE}}$ for quantifying fluid rotation and stretching while requiring much less structured data than LAVD and FTLE.


\begin{figure}
\includegraphics[width=0.9\textwidth]{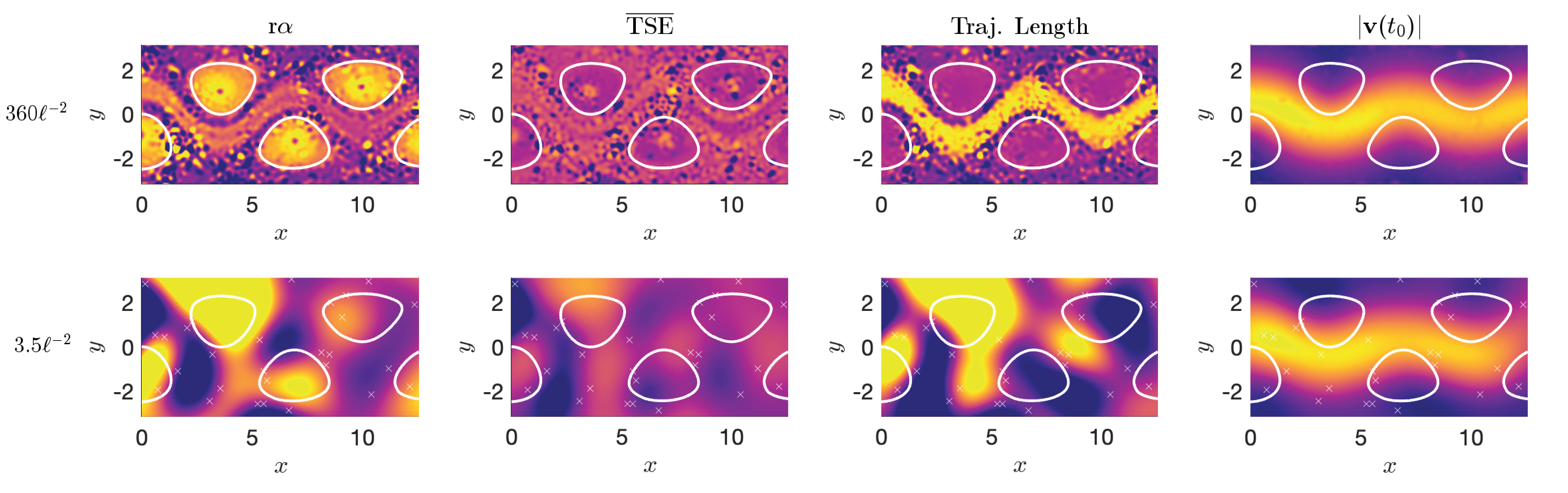}
\caption{Comparison of relative rotation ($\mathrm{r\alpha}$) and stretching ($\mathrm{r\overline{TSE}}$) with non-objective trajectory length and initial particle speed diagnostics. Each diagnostic was progressively and randomly downsampled with the full resolution product created using radial basis functions. For each panel, we include the outermost closed convex LAVD contours to highlight relative agreement of each sparse diagnostic with ground truth material rotation.}
\label{Fig: Bickley Degrade}
\end{figure}

We now progressively and randomly downsample the number of trajectories to test the robustness of our objective diagnostics at low trajectory densities. In figure \ref{Fig: Bickley Degrade}, we calculate $\mathrm{r\alpha}$ and $\rTSE$ as well as the particle trajectory length \citep{Mancho2013} and initial particle speed $|\mathbf{v}(t_0)|$. While trajectory length and particle speed are not objective, they are either easily computed or commonly used diagnostics for experimental LPT studies \citep[see, e.g.,][]{Fu2015, Tauro2017, Rosi2018}. After calculating each diagnostic, we reconstruct the full resolution diagnostic field using radial basis functions, similar to \citet{Encinas-Bartos2022}. Additionally, we include the outermost closed convex LAVD contours from figure \ref{Fig: Bickley Stretch} as a visual reference to aid in the comparisons. 

At a trajectory concentration of 360$\ell^{-2}$, $\mathrm{r\alpha}$ looks relatively similar to its full resolution sampling, but with added noise. Five concentrations of high $\ralpha$ are still distinguishable from the surrounding flow. The thin hyperbolic structures identified by $\rTSE$ in the Bickley jet, however, are already beginning to disappear. This degradation of hyperbolic coherent structures $\rTSE$ is similar to the findings of \citet{Haller2021} for $\overline{\mathrm{TSE}}$ and \citet{Mowlavi2022} for $\overline{\mathrm{TSE}}$ and FTLE. The trajectory length reveals fluid particles in the central jet have the longest trajectories over this time window when compared with nearby particles in the adjacent eddies. The initial velocity magnitude $|\mathbf{v}|$ also appears largest in the center of flow domain, and suggests some oscillating feature may be present.

Dropping the trajectory resolution by a factor of 100 to 3.5$\ell^{-2}$, we again reconstruct the original diagnostic resolution and include the location of trajectory positions with white x's. At this concentration, $\ralpha$ still suggests five distinct local rotation maxima corresponding to the locations of the five eddies. The interpolated $\rTSE$ field no longer recreates any of the features seen at full resolution and hence provides no meaningful information. The trajectory length is also contradicting features that were revealed with 100 times the number of trajectories. At this resolution, we are testing both the ability of the metric to provide useful flow descriptors, as well as the interpolation scheme. We believe, however, that there is limited influence from the interpolation scheme because $|\mathbf{v}|$ is practically constant across all resolutions. We further test the impact of interpolation schemes in Appendix \ref{Appendix: Interpolant}, and refer the reader to \citet{Encinas-Bartos2022} for a deeper investigation of those schemes in a similar context. This trajectory concentration is well below the limits of the previous sparse trajectory studies of \citet{Lekien2010} and \citet{Mowlavi2022}.

\begin{figure}
\includegraphics[width=13cm]{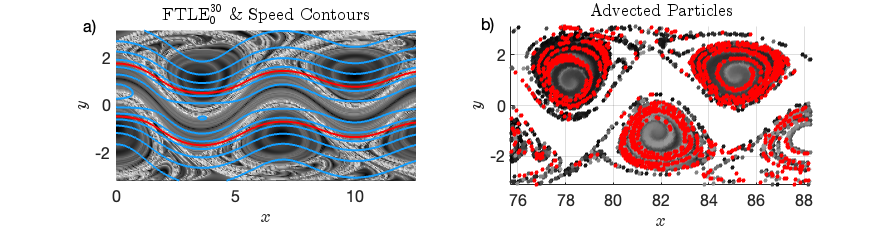}
\caption{Comparison of particle speed, FTLE, and the evolution of fluid particles advected from inside a given velocity level set.}
\label{Fig: Velocity Evolution Bickley}
\end{figure}

From this initial investigation it appears that $|\mathbf{v}|$ may be the best option for identifying structures with low-resolution data. The dominant flow feature suggested by $|\mathbf{v}|$, however, is in fact misleading, even at full resolution. Figure \ref{Fig: Velocity Evolution Bickley}a shows an overlay of $|\mathbf{v}|$ contours with the FTLE field in grayscale. The distribution of $|\mathbf{v}|$ suggests two distinct peaks with a distribution minimum, $s_\delta$, separating the central and outer regions. The fast moving central core is highlighted as the region inside the red $|\mathbf{v}|=s_\delta$ contour in \ref{Fig: Velocity Evolution Bickley}a. Examination of figure \ref{Fig: Velocity Evolution Bickley}a show that the jet and eddy structures suggested by the FTLE field have boundaries that are actually transverse to all of the velocity contours. 

In figure \ref{Fig: Velocity Evolution Bickley}b we plot the final position of fluid particles from \emph{only} the velocity core ($|\mathbf{v}|\ge s_\delta$), colored by their $|\mathbf{v}(t_0)|$ value, after 30 days of advection.  Particles corresponding to the red boundary have been advected as well. It is clear that the velocity-based jet is not actually a coherent structure as the proposed feature has been stretched inside and around the eddies, as well as down the eastward jet, with initial particle speed giving no indication of what feature a particle should be attached to.

\begin{figure}
\centerline{\includegraphics[width=13cm]{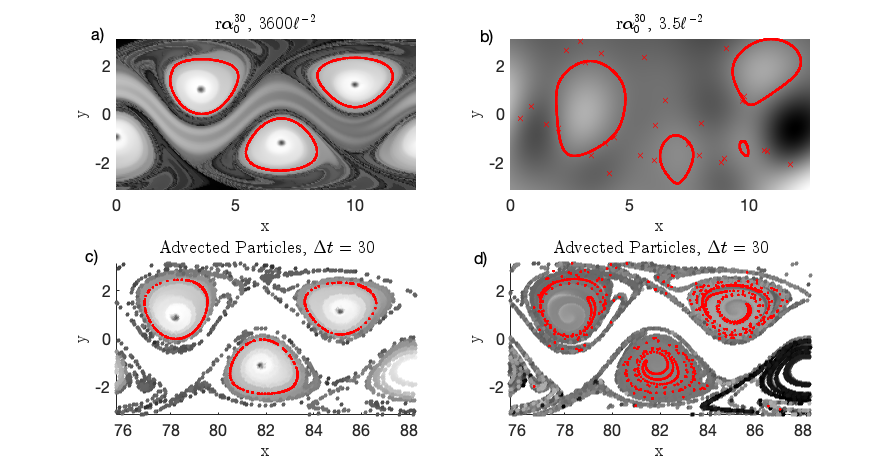}}
\caption{Outermost closed convex $\mathrm{r\alpha}$ level-set contours overlaid on the $\mathrm{r\alpha}$ field. Advected position of $\mathrm{r\alpha}$ contours and surrounding fluid, colored by initial $\mathrm{r\alpha}$ values.}
\label{Fig: rAlpha Advect}
\end{figure}

For comparison, we show the evolution of the $\mathrm{r\alpha}$ structures, those that were similarly resilient to downsampling. In figure \ref{Fig: rAlpha Advect}a we plot $\mathrm{r\alpha}$, with rotationally coherent structure boundaries identified as the outermost closed convex contours of $\mathrm{r\alpha}$, as has been previously used in other elliptic Lagrangian coherent structure methods \citep{Haller2016, Encinas-Bartos2022}. Figure \ref{Fig: rAlpha Advect}c shows the location of the same $\mathrm{r\alpha}$-colored fluid particles after 30 days of advection. The $\mathrm{r\alpha}$ eddy boundaries have minimally deformed after being advected downstream more than 20 times their diameter downstream. This time also corresponds with approximately six full rotations of the eddy, ample time for inaccurate eddy boundaries to experience filamentation. This strong coherence can be attributed to the mathematical definition of $\mathrm{r\alpha}$ as a measure of relative fluid rotation and suggest we have effectively and objectively identified the true rotationally coherent structures in our flow field.

The ability of $\ralpha$ to identify these same eddies at a random lower data concentration ($3.5\ell^{-2}$) is tested in \ref{Fig: rAlpha Advect}b and \ref{Fig: rAlpha Advect}d. We again identify eddies as the outermost closed convex contours of $\ralpha$ at $t=0$ (\ref{Fig: rAlpha Advect}b) and advect them for 30 days. Comparing \ref{Fig: rAlpha Advect}a and \ref{Fig: rAlpha Advect}b, we see the leftmost eddy boundary extending into the central jet and differences arising in the exact eddy boundary location. Testing these boundaries under advection, we find the red boundary particles have remain primarily inside the eddies we sought to identify (\ref{Fig: rAlpha Advect}d), but with a much higher degree of filamentation.

The ability to identify structure boundaries is as much of a test of the interpolation scheme used as of $\ralpha$ to maintain meaningful values at lower resolution. As mentioned above, we investigate interpolation scheme impacts in Appendix \ref{Appendix: Interpolant}. Here, we investigate the ability of low concentration data to provide accurate reference frame values $\overline{\mathbf{v}}$, $\overline{\mathbf{\omega}_e}$, and $\overline{\mathbf{x}}$ and meaningful pointwise $\ralpha$, independent of the interpolation scheme, with the following experiment.

Starting with full resolution ($3600\ell^{-2}$) Bickley jet trajectories, we first calculate ground-truth values for $\ralpha(\mathbf{x}_0)_{\mathrm{full}}$. We then randomly subsample the full resolution trajectory $10^5$ times at a given concentration. For each selection of trajectories, we recalculate $\ralpha(\mathbf{x}_0)_{\mathrm{low}}$ using the lower resolution data, and calculate the correlation coefficient ($R^2$) of $\ralpha(\mathbf{x}_0)_{\mathrm{low}}$ and $\ralpha(\mathbf{x}_0)_{\mathrm{full}}$ at the corresponding locations. We then obtain a distribution of $R^2$ values that details how closely we can approximate the full resolution measurements for a large number of potential trajectory orientations. We then repeat this process at progressively lower concentrations and witness the rate at which the accuracy degrades.

The findings from this experiment are summarized in Table \ref{Table: Bickley Correlation}. The mean value of $R^2$ slowly decreases with increasing sparsity but stays above 0.99, with a standard deviation $\le$0.008 down to $16\ell^{-2}$. The mean of $R^2$ reduces more dramatically around the extremely sparse sampling threshold of $1\ell^{-2}$. This is precisely the trajectory concentration at which we suggest using time-averaged $\boldsymbol{v}_{\mathrm{ref}}$, $\boldsymbol{\omega}_{\mathrm{ref}}$ and $\boldsymbol{x}_{\mathrm{ref}}$ values. The impact of this choice is seen in the final column where mean $R^2$ values nearly return to the $4\ell^{-2}$ value. This suggests a significant improvement in pointwise $\ralpha$ accuracy, and meaningful rotation diagnostics for extremely sparse data.

\begin{table}
\begin{center}
\begin{tabular}{ l c c c c c c c c c} 
 \hline
$\ell^{-2}$:  & 128  & 64 & 32 & 16 & 8 & 4 & 2 & 1 & 1 (ref) \\ [0.5ex] 
 \hline\hline
$\mu$: & 0.997  & 0.997 & 0.996 & 0.991 & 0.98 & 0.95 & 0.87 & 0.68 & 0.93\\
 \hline
$\sigma$: & $2\times10^{-4}$  & 0.001 & 0.003 & 0.008 & 0.018 & 0.045 & 0.11 & 0.25 & 0.14\\
 \hline
\end{tabular}
\caption{Bickley Jet downsampling effect on $\ralpha$ accuracy. Mean ($\mu$) and standard deviation ($\sigma$) of $R^2$ values correlating downsampled $\ralpha$ with full resolution calculations. Raw $\ralpha$ values maintain a strong correlation until the extremely sparse sampling threshold (1 $\ell^{-2})$, at which point steady time-averaged reference values ('ref', Section \ref{Sec: Extreme}) allow for a significant improvement in accuracy again.}
 \label{Table: Bickley Correlation}
\end{center}
\end{table}

\subsubsection{AVISO}
\label{sec:AVISO}
We now consider the identification of elliptic Lagrangian flow structures from a two-dimensional ocean satellite altimetry data provided by AVISO which has been the focus of several coherent structure studies \citep[see, e.g.,][]{Haller2016, Haller2021}.
The zonal and meridional component $ \mathbf{v} = (\mathrm{v}_1, \mathrm{v}_2) $ of the ocean currents are derived from the sea-surface height profile 
\begin{align}
f\mathrm{v}_2(\mathbf{x},t) &= \dfrac{1}{\rho} \dfrac{\partial}{\partial x}p(\mathbf{x},t) \\
f\mathrm{v}_1(\mathbf{x},t) &= \dfrac{1}{\rho} \dfrac{\partial}{\partial y}p(\mathbf{x},t) \\
-g &= \dfrac{1}{\rho}\dfrac{\partial}{\partial z}p(\mathbf{x},t),
\end{align} where $ p(\mathbf{x},t) $ is the pressure, $ \rho $ is the fluid density, $ f $ is the Coriolis parameter and $ g $ is the Earth's acceleration. The daily-gridded velocity data is freely available from the Copernicus Marine Environment Monitoring Service. While this is an observational product, the processed nature of the product prevents many of the complications that we will see with our experimental examples in the following section. Our analysis focuses on the North Atlantic Gulf Stream between longitudes $70^{\circ}$W and $55^{\circ}$W, and latitudes $30^{\circ}$N and $45^{\circ}$N, spanning September and October 2006. We start with an initial grid of trajectories consisting of $ 150 \times 150 $ points. Using a characteristic length scale of mesoscale eddies ($\ell=100$ km), this density roughly corresponds to $ 100\ell^{-2} $.
\begin{figure}
\includegraphics[width=\textwidth]{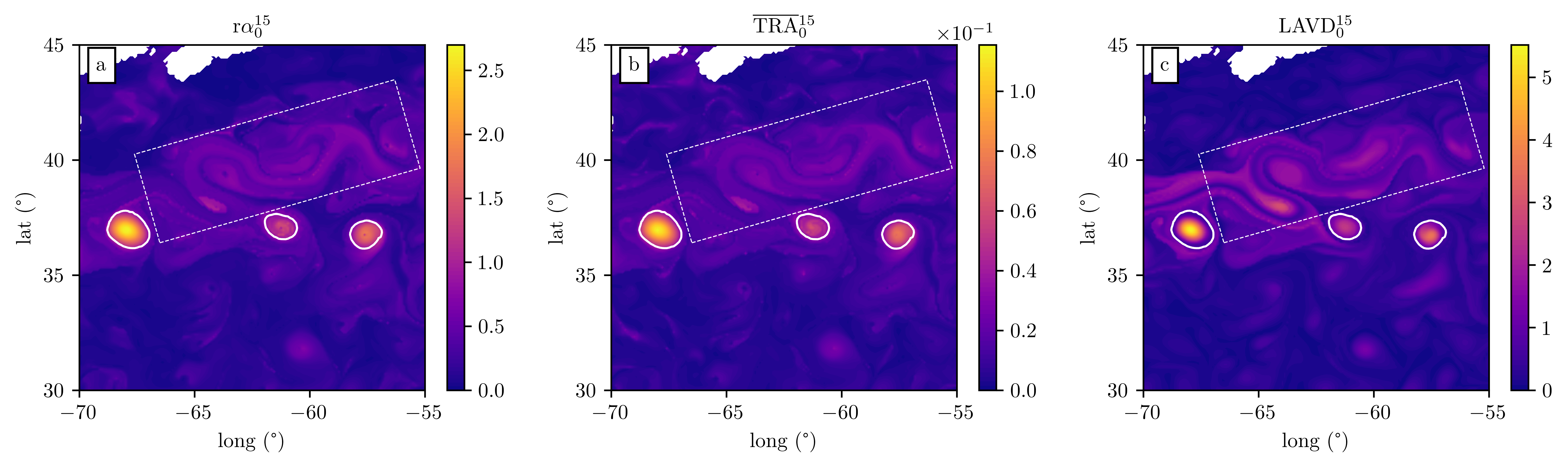}
\caption{Overview of rotation diagnostics for the AVISO data in the Gulf Stream at full resolution ($100 \ell^{-2}$). The white contours are the outermost convex contours of the LAVD-field. The dashed white box highlights the meanders of the Gulf Stream.}
\label{Fig: AVISO_Overview}
\end{figure}

Elsewhere in the Atlantic, \cite{Haller2021} and \cite{Encinas-Bartos2022} have shown that the AVISO velocity field is slowly varying, with a vanishing spatially-averaged vorticity. Furthermore, these authors showed that $ \mathrm{\overline{TRA}} $  effectively identifies ocean eddies, and outperforms other sparse trajectory rotation diagnostics. As such, in figure \ref{Fig: AVISO_Overview} we compare the $ \mathrm{r\alpha} $ (\ref{Fig: AVISO_Overview}a) with $ \mathrm{\overline{TRA}} $ (\ref{Fig: AVISO_Overview}b), using LAVD as the ground truth of material rotation (\ref{Fig: AVISO_Overview}c). The white dashed box highlights the meanders of the Gulf Stream. Three eddy boundaries are identified as outmost closed convex contours of LAVD in \ref{Fig: AVISO_Overview}a-c.

When compared with LAVD, $ \mathrm{\overline{TRA}} $ faithfully approximates the LAVD field, identifying the meanders of the Gulf Stream, high rotation in the adjacent eddies, and relatively low rotation away from these central features. At this trajectory density, figure \ref{Fig: AVISO_Overview}a reveals that $ \mathrm{r\alpha} $ can match the ability of $ \mathrm{\overline{TRA}}$ to visualize the correct flow structure topology. This stems from the similarity in $ \mathrm{\overline{TRA}}$ and $\ralpha$ definitions for small bulk advection and rotation, as highlighted in Section \ref{Sec: Extreme}.

\begin{figure}
\includegraphics[width=14cm]{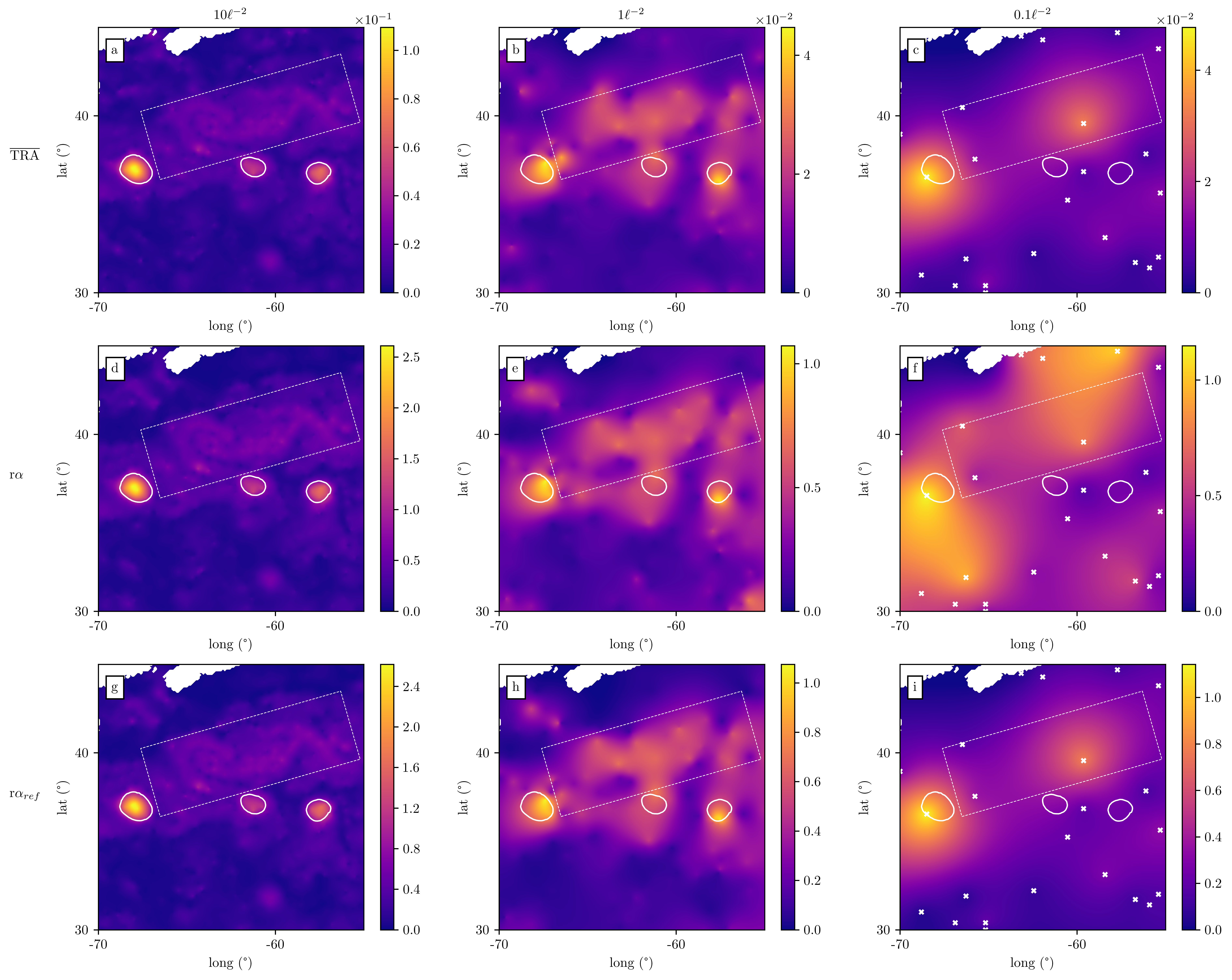}
\caption{Comparison for the AVISO data between objective rotational sparse data-diagnostic $ \mathrm{r\alpha}$, $\mathrm{r\alpha_{\mathrm{ref}}} $ and its quasi-objective single trajectory counterpart $ \mathrm{\overline{TRA}} $ under varying trajectory densities ($10 \ell^{-2}, 1 \ell^{-2}, 0.1 \ell^{-2}$).  The white markers denote the trajectory endpoints.}
\label{Fig: AVISO_Overview_Subsampling}
\end{figure}

We compare the sparse data performance of $\mathrm{r\alpha}$ against $\mathrm{\overline{TRA}}$ in figure \ref{Fig: AVISO_Overview_Subsampling} by progressively and randomly subsampling using the same methodology as in section \ref{Section: Bickley}. We again interpolate back to full resolution using a radial basis function for each successive subsampling. We also include the major eddy boundaries from Fig. \ref{Fig: AVISO_Overview}.

At a trajectory concentration of $10\ell^{-2}$, $\mathrm{\overline{TRA}}$ (\ref{Fig: AVISO_Overview_Subsampling}a) and $\mathrm{r\alpha}$ (\ref{Fig: AVISO_Overview_Subsampling}d) appear qualitatively similar, with both suggesting relatively high rotation in the Gulf Stream meanders, and three distinct elliptic rotational maxima aligned with the eddy boundaries. For a trajectory density of $1 \ell^{-2}$, $ \mathrm{r\alpha}$ (\ref{Fig: AVISO_Overview_Subsampling}e)  is qualitatively indistinguishable from $ \mathrm{\overline{TRA}} $ (\ref{Fig: AVISO_Overview_Subsampling}b) with two of the three mesoscale LAVD eddies still visible as local maxima in both diagnostic fields. For this region of the ocean, we often have ocean drifter observations (e.g., GDP) at concentrations between $1 \ell^{-2}$ and $0.1 \ell^{-2}$, as will be discussed in Section \ref{sec:OceanDrifters}.

As we pass the extremely sparse sampling threshold to $0.1 \ell^{-2}$, the leftmost eddy is still identifiable as a $\mathrm{\overline{TRA}}$ and $ \mathrm{r\alpha}$ maximum (Fig. \ref{Fig: AVISO_Overview_Subsampling}c, f, respectively) due to the presence of the single trajectory originating inside its boundary (white x). The remaining eddies are no longer visible for either metric due to a lack of data, but $\mathrm{\overline{TRA}}$ still highlights a relatively higher rotation in the Gulf Stream. The local maximum $\mathrm{\overline{TRA}}$ within the dashed white box may be incorrectly classified as an eddy, but it actually corresponds to a region of high material rotation induced by shear in the Gulf Stream. In contrast, $\ralpha$ has created multiple local maxima near the coast of Nova Scotia not present at higher concentrations.

At this extremely sparse resolution, we again replace $ \boldsymbol{\overline{\omega}}(t) $,  $ \boldsymbol{\overline{v}}(t) $, and $ \boldsymbol{\overline{x}}(t) $  with their time-averaged steady approximations, $ \boldsymbol{v}_{\mathrm{ref}}, \boldsymbol{\omega}_{\mathrm{ref}}, \boldsymbol{x}_{\mathrm{ref}}$ from equations (\ref{Eq: Time Avg v}-\ref{Eq: Time Avg x}). As with the Bickley Jet statistics, this improves the accuracy of $\ralpha$ with $\ralpha_{\mathrm{ref}}$ displayed in \ref{Fig: AVISO_Overview_Subsampling}i. With the steady reference frame, $\ralpha_{\mathrm{ref}}$ and $\mathrm{\overline{TRA}}$ again show nearly indistinguishable flow topologies. The coastal local maxima disappear, and the local maximum appears in the shear dominated meanders of the Gulf Stream.  Furthermore, if we apply this steady reference frame change to higher concentrations (Figs. \ref{Fig: AVISO_Overview_Subsampling}g-h), we obtain flow structure visualizations comparable to $\ralpha$. This suggests that obtaining meaningful results using the extremely sparse steady reference-frame approach is resilient over a range of concentrations.

\begin{figure}
\includegraphics[width=14cm]{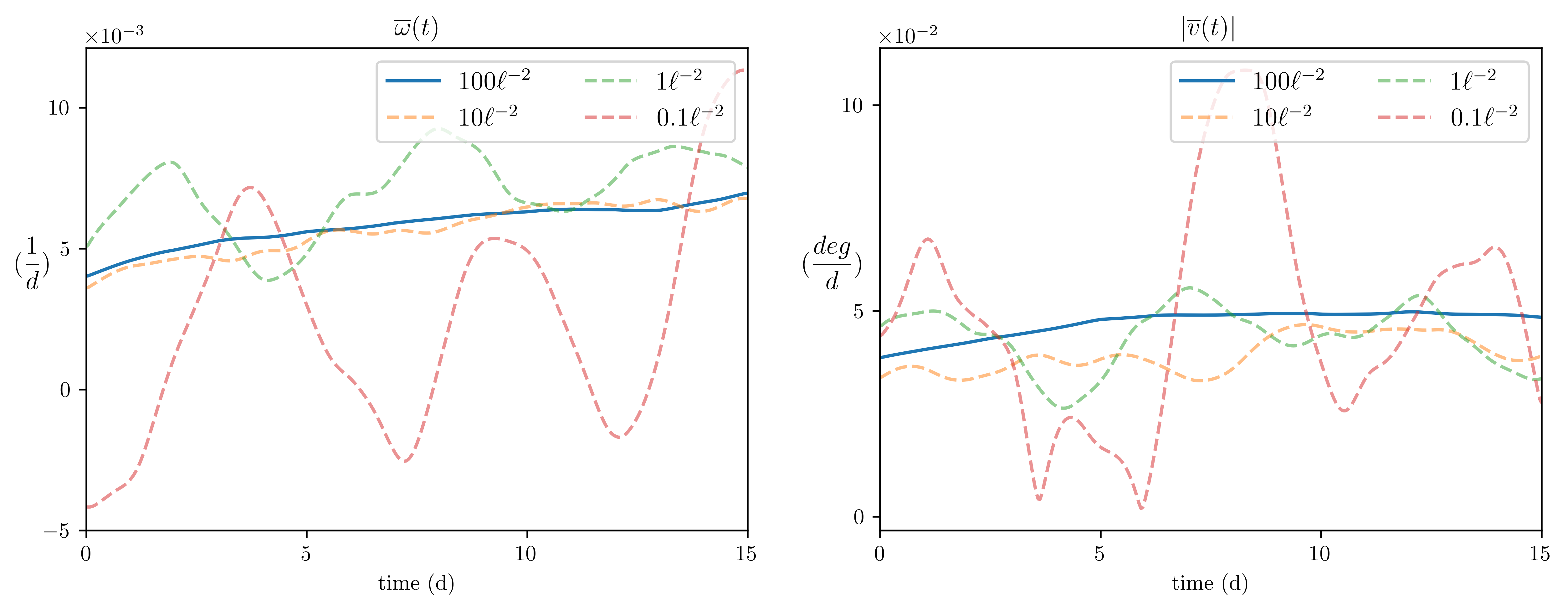}
\caption{Data-based estimates for bulk translation velocity $ |\boldsymbol{\overline{v}}(t)| $ and rotation $ \boldsymbol{\overline{\omega}}(t) $ under varying trajectory density ($100 \ell^{-2}, 10 \ell^{-2}, 1 \ell^{-2}, 0.1 \ell^{-2}$).}
\label{Fig: AVISO_MeanProperties}
\end{figure}

This effect stems from the impact of undersampling on bulk rotation and translation estimates. In Figure \ref{Fig: AVISO_MeanProperties}, we display the $ \boldsymbol{\overline{\omega}}(t) $ and $ \boldsymbol{\overline{v}}(t)$ estimates for each trajectory concentration. At $ 100 \ell^{-2} $ the mean flow properties are nearly constant over 15 days. This is consistent with the expectation from earlier studies that Eulerian features do not change significantly over such relatively small time scales in the ocean. As we progressively reduce the trajectory density, however, $ \boldsymbol{\overline{\omega}}(t) $ and $ \boldsymbol{\overline{v}}(t) $ start displaying strong oscillations in time. At extremely low trajectory densities ($ <1 \ell^{-2} $), the spatial averages are no longer representative of the underlying mean flow properties (dashed red curves in Figure \ref{Fig: AVISO_MeanProperties}).  These oscillations contribute unphysical fluctuations in the direction of $ \boldsymbol{v}_{d}$ and hinder our ability to measure rotation with $ \mathrm{r\alpha} $.

To further quantify the impact of sparse sampling on $\ralpha$, we repeat the $R^2$ distribution analysis that was performed for the unsteady Bickley jet. Using AVISO surface current data to generate an initial baseline at $100\ell^{-2}$, we run $10^4$ random subsamplings at a range of concentration levels. The summary statistics of this experiment can be found in Table \ref{Table: AVISO Correlation}. A consistent decrease in the $R^2$ mean, and increase in $R^2$ standard deviation is found as the data concentration levels decrease, though the level of $\ralpha$ accuracy remains quite high. Even at a low concentration level of $1\ell^{-2}$, we obtain a mean $R^2$ value of 0.92. 

\begin{table}
\begin{center}
\begin{tabular}{ l c c c c c c c c c c} 
 \hline
$\ell^{-2}$:  & 64 & 32 & 16 & 8 & 4 & 2 & 1 & 0.1 & 1 (ref) & 0.1 (ref) \\ [0.5ex] 
 \hline\hline
$\mu$: & 0.999 & 0.998 & 0.996 & 0.992 & 0.98 & 0.96 & 0.92 & 0.45 &0.95 & 0.89 \\
 \hline

$\sigma$: & $2\times10^{-4}$  & $7\times10^{-4}$ & 0.002 & 0.003 & 0.008 & 0.02 & 0.04 &0.21 & 0.02  & 0.13 \\
 \hline
\end{tabular}
\caption{Downsampling effect on $\ralpha$ accuracy for AVISO Gulf Stream surface currents. Mean and standard deviation of $R^2$ values relating downsampled $\ralpha$ with full resolution calculations for $10^4$ random subsamplings at each resolution. Raw $\ralpha$ values maintain a strong correlation until the extremely sparse threshold $(1 \ell^{-2})$, at which point time-averaged reference values (ref) allow for a significant improvement in accuracy again.}
 \label{Table: AVISO Correlation}
\end{center}
\end{table}

The most precipitous drop in $\ralpha$ accuracy is found from $1\ell^{-2}$ to $0.1\ell^{-2}$. One such example of this poor performance was highlighted in Figure \ref{Fig: AVISO_Overview_Subsampling}. We further validate the improvement in $\ralpha$ accuracy at this extreme sparsity level in the last two columns of Table \ref{Table: AVISO Correlation}. After transitioning to time-averaged reference frame values, there is a significant increase in accuracy, with the mean of $R^2$ jumping from 0.92 to 0.95 for $1\ell^{-2}$, and from 0.45 to 0.89 for $0.1\ell^{-2}$. With this simple reference-frame methodology we find r$\alpha$ is a promising Lagrangian flow rotation diagnostic for extremely sparse data.

\subsubsection{Stationary Concentrated Vortex}

Our third example considers the identification of elliptic coherent structures inside the stationary concentrated vortex model (SCVM) of \citet{Onishchenko2021}. The SCVM represents finite-size cyclones in the Earth's atmosphere as steady axially-symmetric solutions to the Euler equations. Both the vertical and radial extent of the vortex is controlled by model parameters. The SCVM fundamentally differs from the previous flows as it is a highly rotational flow with no bulk translation. Our investigation of the SCVM helps confirm that $\ralpha$ and $\rTSE$ are not tailored to a specific kind of flow, and can still outperform standard Eulerian diagnostics even when there are no transient features in the flow.

The SCVM consists of inner and outer regions, an internal upward motion and external downward motion. A distinct central torus was also identified by \citet{Aksamit2023b} as a Lagrangian coherent structure. The cyclone is further divided vertically into a centripetal and centrifugal flow with upward moving fluid recirculated from the top of the vortex back down. 

The SCVM was originally derived as an axisymmetric flow from a streamfunction $\Psi(r,v)$ and can be written in cylindrical coordinates as

\begin{align*}
\mathrm{v}_r & = -\mathrm{v}_0 \frac{r}{L} \left( 1-\frac{z}{L} \right)e^{-\frac{z}{L}-\frac{r^2}{r_0^2}} \\
\mathrm{v}_\theta & = \pm \mathrm{v}_{\theta_0}\frac{z}{L}e^{-\frac{z}{L}-\frac{r^2}{r_0^2}} \\
\mathrm{v}_z & = 2\mathrm{v}_0 \frac{z}{L} \left( 1-\frac{r^2}{r_0^2} \right)e^{-\frac{z}{L}-\frac{r^2}{r_0^2}},
\end{align*}
where $r_0$ is the radial extent separating inner and outer vortex structures, $L$ is the height of maximum vertical velocity at $r=0$, and $\mathrm{v}_0$ and $\mathrm{v}_{\theta_0}$ are characteristic velocities. For our example, we use $\mathrm{v}_0 = 1$, $L = 10$, $r_0 = 1 $ and $\mathrm{v}_{\theta_0}=9$. We consider our characteristic length of the flow to be equal to the separation of inner and outer motions, $\ell=r_0=1$, and the approximate radius of the main central torus.

\begin{figure}
\includegraphics[width=\textwidth]{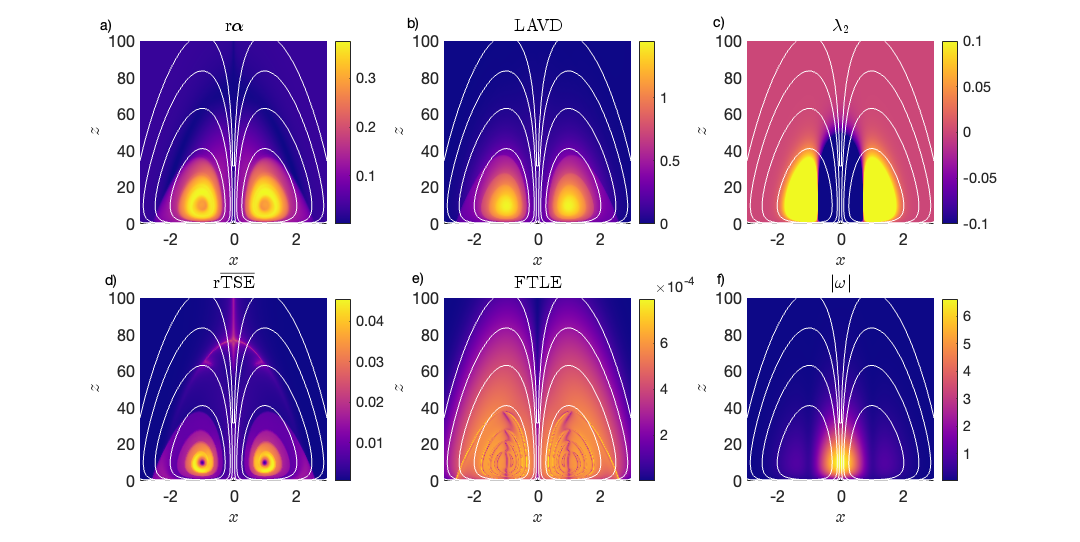}
\caption{Comparison of rotation diagnostics $\mathrm{r\alpha}$, $\mathrm{LAVD}$, $\lambda_2$, and $|\boldsymbol{\omega}|$ for the stationary concentrated vortex model in the $y=0$ plane. White contour lines are contours of the SCVM stream function $\Psi$, indicating circulation patterns in the model. Note the agreement between $\mathrm{r\alpha}$, $\mathrm{LAVD}$, $r\mathrm{\overline{TSE}}$ and $\mathrm{FTLE}$ features with the stream function structures, and the dissimilarity of $|\boldsymbol{\omega}|$, and $\lambda_2$ features.}
\label{Fig: SCVM Full Res}
\end{figure}

In figure \ref{Fig: SCVM Full Res} we compare the objective rotational diagnostics $\mathrm{r\alpha}$ and $\mathrm{LAVD}$ advected for $\Delta t = 500$ with the vorticity magnitude, $|\boldsymbol{\omega}|$ and the widely used lambda-2 criterion of \citet{Jeong1995}. If $\mathbf{J}$ is the velocity gradient tensor, and $\mathbf{S}$ and $\mathbf{\Omega}$ are the symmetric and antisymmetric parts, the lambda-2 criterion of \citet{Jeong1995} suggests that regions with negative intermediate eigenvalues of $\mathbf{S}^2 + \mathbf{\Omega}^2$ (labeled $\lambda_2$) are part of a vortex core. Stationarity of the flow suggests $\overline{\mathrm{TRA}}$ would also be a suitable comparison, but in light of the strong performance of r$\alpha$ against $\overline{\mathrm{TRA}}$ in the previous two examples, we have left it out for compactness of presentation. 

From initial high spatial resolution ($2000\ell^{-3}$) calculations, a two-dimensional slice along the $y=0$ plane is shown for each field in figure \ref{Fig: SCVM Full Res}. We overlay these plots with the level-set contours of the SCVM streamfunction $\Psi(r,v)$. As this is a stationary axisymmetric flow, $\Psi$-contours provide us a ground truth of transport barrier structure. A qualitative contour comparison of the sparse metrics ($\ralpha$ and $\rTSE$) with $\Psi$ show a comparable level of structure agreement as one can find from LAVD and FTLE structures. We find that $\mathrm{r\alpha}$ and $\mathrm{LAVD}$ contours suggest very similar rotational behavior, with both indicating the same location of a central torus vortex surrounded by recirculating flow. r$\overline{\mathrm{TSE}}$ and FTLE show strong spiraling into the center of the vortices identified by $\mathrm{r\alpha}$ and $\mathrm{LAVD}$. One notable difference between sparse and spatial-gradient-reliant diagnostics exists in the center of the toroidal vortex core in $\ralpha$ and $\rTSE$ which again have unnaturally low values when compared with LAVD and FTLE, as we also saw in the Unsteady Bickley jet. 

We see a stark difference, however, when comparing $\Psi$ with $|\boldsymbol{\omega}|$ and $\lambda_2$ fields. Contours of $|\boldsymbol{\omega}|$ and $\lambda_2$ are often orthogonal to $\Psi$-contours, suggesting a disagreement in structure identification. We further investigate these qualitative findings by testing the ability of our rotation diagnostics to identify the dominant elliptic coherent structures in this vortex flow. Potential vortex boundaries are identified as the set of outermost closed convex contours for $\ralpha$, LAVD, and $|\boldsymbol{\omega}|$. Following the derivation of the $\lambda_2$ criterion, we use $\lambda_2<0$ as the designation of strong rotational motion \citep{Jeong1995}. Fluid parcels from inside the boundaries of these features are then advected for twice the integration time used for the Lagrangian diagnostics, $\Delta t = 1000$. This approach allows us to test both the Lagrangian and Eulerian diagnostics to identify features that maintain their coherence beyond the time horizon of data used for the initial boundary calculation.

\begin{figure}
\centerline{\includegraphics[width=10cm]{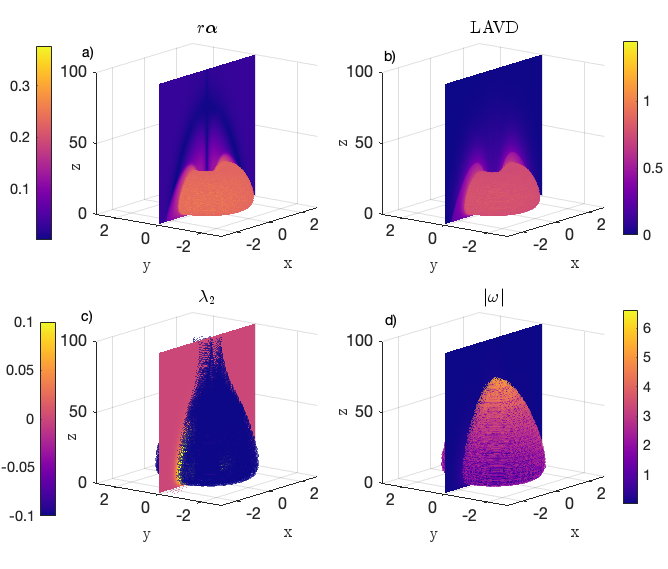}}
\caption{Trajectories corresponding to coherent structures identified by the four diagnostics, calculated for the same integration times}
\label{Fig: SCVM Tracks}
\end{figure}

These fluid particle paths are shown in figure \ref{Fig: SCVM Tracks} with trajectories colored by the initial diagnostic field value. We include the corresponding figure \ref{Fig: SCVM Full Res} diagnostic field on the $y=0$ plane as a reference. The vortex core identified by $\mathrm{r\alpha}$ and $\mathrm{LAVD}$ maintain coherence over time, with seemingly no deviation of fluid from the domain identified as the central torus vortex (Figure \ref{Fig: SCVM Tracks}a-b). Given that $\mathrm{r\alpha}$ does not rely on the same three-dimensional velocity gradient data as $\mathrm{LAVD}$, $\mathrm{r\alpha}$ shows exceptional performance in identifying this Lagrangian coherent structure. Fluid initially inside $\lambda_2$ and $|\boldsymbol{\omega}|$ structures have instead maintained no coherence, and are sufficiently dispersed in the flow. The high $|\boldsymbol{\omega}|$ (Fig. \ref{Fig: SCVM Tracks}d) fluid parcels are now mixed throughout the entire inner SCVM ``beehive" domain, inhabiting regions with both high and low $|\boldsymbol{\omega}|$ values. The $\lambda_2<0$ fluid is recirculating in both the inner and outer regions of the SVCM (Fig. \ref{Fig: SCVM Tracks}c), with the $\lambda_2$-estimated vortex core parcels inhabiting the domain of both negative and positive $\lambda_2$ values. This is distinct from LAVD and $\mathrm{r\alpha}$ vortex trajectories that stay confined to the regions of high LAVD and $\mathrm{r\alpha}$, respectively.

\begin{figure}
\includegraphics[width=\textwidth]{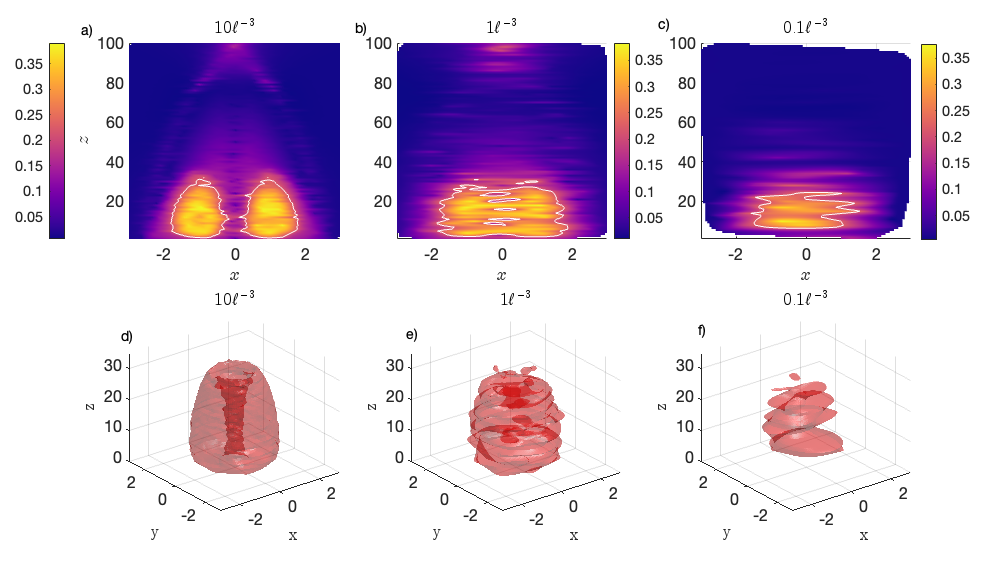}
\caption{Top Row: $\mathrm{r\alpha}_0^{500}$ on the $y=0$ plane reconstructed from randomly selected trajectories at decreasing particle density. The inner vortex is extracted at each concentration as the closed contours corresponding to the level-set value from Figure \ref{Fig: SCVM Full Res}. Bottom Row: Comparison of the inner vortex surface in $\mathrm{r\alpha}_0^{500}$ fields reconstructed from randomly selected trajectories at decreasing particle density. The inner vortex is extracted in each scenario as closed contours corresponding to the outermost closed convex contour in Figure \ref{Fig: SCVM Tracks}. }
\label{Fig: SVCM Plane Degrade}
\end{figure}

The ability of $\mathrm{r\alpha}$ to identify these same Lagrangian coherent structures at increasing sparsity is shown in Figure \ref{Fig: SVCM Plane Degrade}. As our only three-dimensional numerical example, we utilize a random set of initial particle trajectories in three dimensions and extract isosurfaces from a reconstructed $\mathrm{r\alpha}$ field using natural neighbor interpolation. We utilize the same $\mathrm{r\alpha}$ value for 3D level-set extraction as the outermost closed convex contour used to test vortex boundaries (figure \ref{Fig: SCVM Tracks}). Plots of reconstructed $\mathrm{r\alpha}$ fields on the $y=0$ plane (figure \ref{Fig: SVCM Plane Degrade}a-c) reveal that small deviations from this range would not significantly change the topology of the surfaces in figure \ref{Fig: SVCM Plane Degrade}d-f. Note the difference of $z$-value ranges between figure \ref{Fig: SVCM Plane Degrade}a-c and figure \ref{Fig: SVCM Plane Degrade}d-f.

In figure \ref{Fig: SVCM Plane Degrade}, $\mathrm{r\alpha}$ at $10\ell^{-3}$ looks only moderately different from $2000\ell^{-3}$. Two distinct intersections of the central torus with the $y=0$ plane are still evident in white, and the annular shape is clearly visible in the associated isosurface. At $1\ell^{-3}$, the "donut hole" above $x=0$ becomes less obvious, but the horizontal and vertical extent of the central torus is still  accurate. For $0.1\ell^{-3}$, we use eqs (\ref{Eq: Time Avg v}-\ref{Eq: Time Avg w}) to implement the same bulk time-averaged reference values for $ \boldsymbol{v}_{\mathrm{ref}}, \boldsymbol{\omega}_{\mathrm{ref}}, \boldsymbol{x}_{\mathrm{ref}} $. At this extremely sparse sampling, the correct extent of strong central rotation is still evident, as is a suggestion of the inner beehive feature, but the vortex surface is now sufficiently deformed that it no long resembles a torus.

\subsection{Experimental Data}

In the previous section we investigated the ability of $\mathrm{r\alpha}$ and $r\overline{\mathrm{TSE}}$ to accurately represent elliptic and hyperbolic Lagrangian coherent structures at increasing sparsity in numerical data. In this section, we show how the sparse capabilities of $\mathrm{r\alpha}$ apply to real experimental data for a two-dimensional and three-dimensional flow.

\subsubsection{Global Drifter Database}
\label{sec:OceanDrifters}
For our first experimental data set we consider a set of drifters from the Global Drifter Program (GDP) \citep{Lumpkin_GDP2019} located in the North Atlantic Ocean. The GDP data set contains more than 40000 drifters spanning the last 4 decades. Roughly 1200 drifters are currently active worldwide and report their position every 6 hours \citep{Lumpkin2007}. 

\begin{figure}
\centering \includegraphics[width=\textwidth]{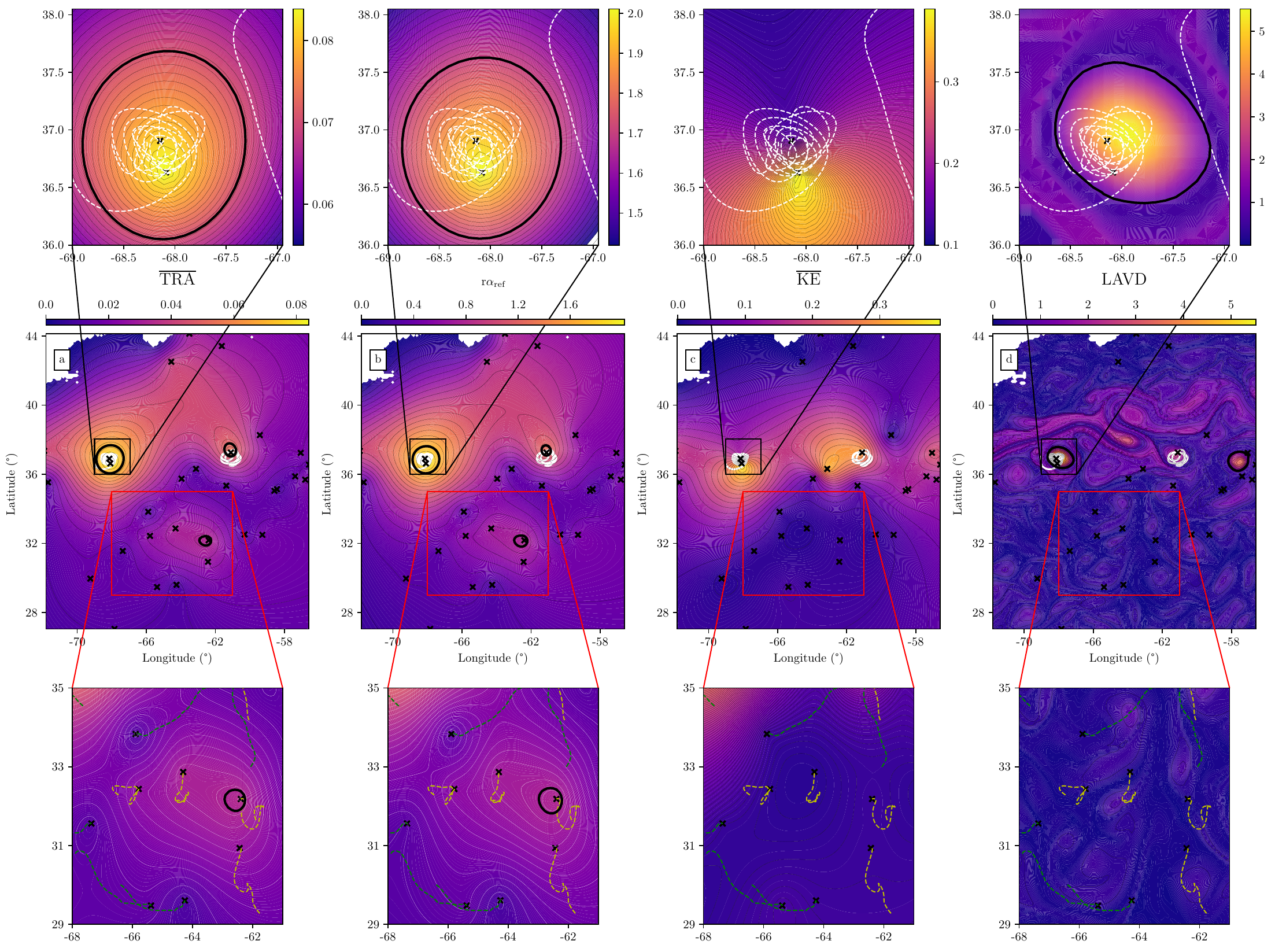}
\caption{Drifter based $ \overline{\mathrm{TRA}} $ (a), $ \mathrm{r\alpha} $ (b), $ \mathrm{\overline{KE}} $ (c), and LAVD (d) in the North Atlantic Gulf Stream. The black x's indicate drifter positions.}
\label{fig:GDP}
\end{figure}

Specifically, we focus here on a subset of drifters active over 15 days from 20 September to 4 October 2006 in the Gulf Stream (Figure \ref{fig:GDP}) during the same time period of analysis as in section \ref{sec:AVISO}. In figure \ref{fig:GDP}, we reconstruct diagnostic fields from the true GDP tracks with a drifter density of $ 0.63 \ell^{-2} $, for $ \ell = 100 \mathrm{km} $. We are thus in an extremely sparse data setting ($ < 1 \ell^{-2} $) and we again compute $ \mathrm{r\alpha} $ by replacing the time-dependent mean flow properties with time-averaged $ \boldsymbol{\omega}_{\mathrm{ref}}, \boldsymbol{v}_{\mathrm{ref}} $ and $ \boldsymbol{x}_{\mathrm{ref}} $ from the drifter data.

Following the sparse eddy identification study by \citet{Encinas-Bartos2022}, we compare two sparse drifter-based rotation diagnostics, $ \mathrm{r\alpha_{\mathrm{ref}}}$ in \ref{fig:GDP}a and $ \overline{\mathrm{TRA}} $ in \ref{fig:GDP}b, with the trajectory-averaged kinetic energy (KE, figure \ref{fig:GDP}c), and the full resolution LAVD field using AVISO data from section \ref{sec:AVISO}. Though the AVISO current data may not be representative of the full spatial scale of flow features affecting the drifters, we include LAVD as a mesoscale comparison for general flow behavior. Additionally, we extract looping segments from the drifter trajectories as proposed by the frame-dependent algorithm from \cite{Lumpkin2016}. The white trajectory segments indicate anticyclonic loopers which frequently arise in connection with ocean eddies \citep[see, e.g.,][]{Griffa2008, Dong2011}. The closed black curves in Figures \ref{fig:GDP}a, b and d denote the approximate vortex boundaries respectively obtained from drifter-based $ \overline{\mathrm{TRA}} $ and $\ralpha_{ref}$, and AVISO-based LAVD, respectively .

Both $ \mathrm{r\alpha_{\mathrm{ref}}} $ and $ \overline{\mathrm{TRA}} $ are able to detect rotational features just to the south of the Gulf Stream from the extremely sparse GDP data. Anticyclonic loopers are also observed in the regions surrounding the principal local maxima of $ \mathrm{r\alpha_{\mathrm{ref}}} $ and $ \overline{\mathrm{TRA}} $ . The leftmost eddy contains two looping drifters that display high rotational values (figure \ref{fig:GDP}a-b). In contrast, the $ \mathrm{\overline{KE}} $ fails to clearly identify the leftmost eddy since it displays sharp gradients between the two drifters inside the eddy (figure \ref{fig:GDP}c). Both $ \mathrm{r\alpha_{\mathrm{ref}}} $ and $ \mathrm{\overline{TRA}} $ display a weaker rotational feature at roughly $32^{\circ}$N latitude and $ 64^{\circ} $W longitude. No loopers are identified in this region, but the yellow trajectories in the high $ \mathrm{r\alpha} $ and $ \overline{\mathrm{TRA}} $ zones show greater circulating behavior with respect to the surrounding green trajectories (figure \ref{fig:GDP}, bottom insets), suggesting stronger rotational motion in the surface currents. There is no feature that resembles any vortical structure in the $ \mathrm{\overline{KE}} $ field in this region. 

In this ocean buoy experiment, we were able to identify eddies adjacent to the Gulf Stream in both $ \mathrm{\overline{TRA}} $ and $ \mathrm{r\alpha} $ fields that were not identified using existing sparse data diagnostics. In contrast with the $ \mathrm{\overline{TRA}} $, which requires additional assumptions on the frame of reference, the $ \mathrm{r\alpha} $ returns structures that are valid for all observers. This suggests that $\ralpha$ may be beneficial for systematic eddy analysis of the GDP dataset that can complement remote-sensing based tools that lack submesoscale motion resolutions. Additionally, $\mathrm{r\alpha}$ shows great promise as a buoy diagnostic for focused buoy deployments (e.g., LASER, GLAD, etc.) where buoy concentrations higher than $1\ell^{-2}$ will result in more refined descriptions of the flow.

\subsubsection{Large-scale LPT Experiment}

Our last example involves a novel, large-volume three-dimensional Lagrangian particle tracking velocimetry dataset from an industrial-scale wind tunnel \citep{Hou2021}. Utilizing glare-point spacing on large soap bubbles, \citet{Hou2021} were able to investigate the vortical wake behind a tractor-trailer model at a $9^{\circ}$ yaw angle. For details on the methodology, see \citet{Kaiser2023}. We compare two wind-tunnel experiments, one with the tractor-trailer model obstructing the flow, and a uniform flow without the tractor-trailer.

Each experiment provides around 300 $s$ of trajectory data, sampled at 150 Hz with an average windspeed of 10 ${m s^{-1}}$. The bubble generator and camera setup produce time-varying particle concentrations at the very lower end of our computation spectrum. For measurements of the tractor-trailer wake, trajectory concentrations oscillate between 0.2 and 1.5$\ell^{-3}$ with a mean concentration of 0.6$\ell^{-3}$. The uniform flow measurements have a slightly higher resolution, ranging from 0.5 to 3$\ell^{-3}$ and a mean concentration of 1.3$\ell^{-3}$.  To avoid significant oscillations in relative Lagrangian velocities, we again utilize the time-averaged reference frame determined by the bubble trajectories. Furthermore, as we are using trajectory data that quickly transits through the observation volume, we investigate 10-frame increments and require all trajectories to be present for every frame. At these extremely low concentrations, a single particle that disappears and reappears during the integration window may have a substantial impact on the deformation velocity of other fluid particles. A 10-frame integration time was chosen as a balance between the length of integration time and the number of tracks available to calculate meaningful mean values.

\begin{figure}
\includegraphics[width=0.8\textwidth]{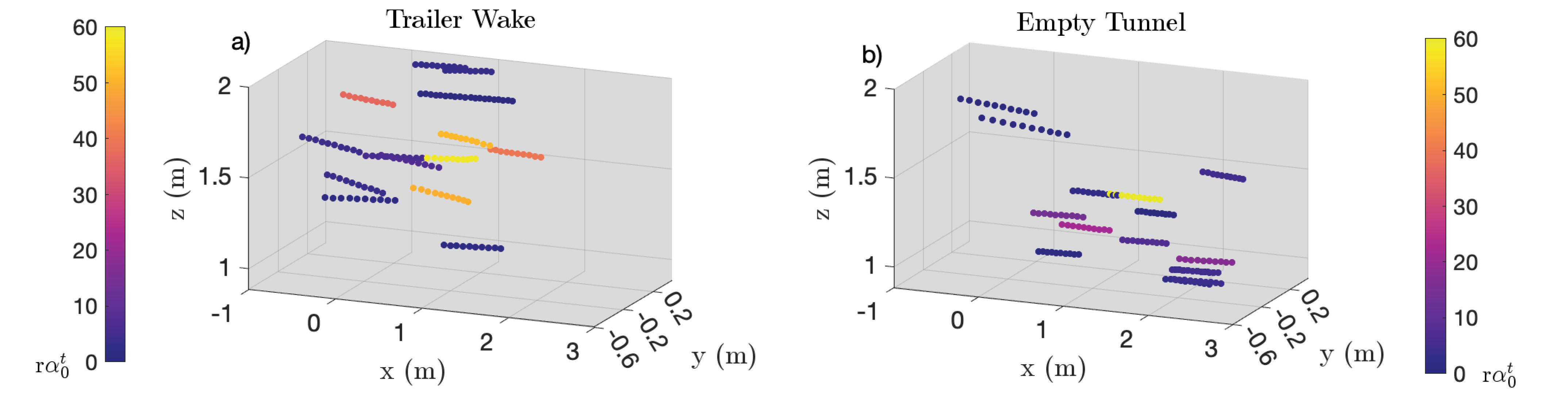}
\caption{An example of one randomly selected 10-frame (approximately 0.07 s) instance of bubble tracks colored by their $\ralpha_0^t$ in the two LPT experiments. At these observation concentrations, it is difficult to identify coherent flow features, and impossible to calculate meaningful spatial derivatives. Accumulated statistics of $\ralpha_0^t$ can start to differentiate behavior in the two flows.}
\label{Fig: Bubble Tracks}
\end{figure}

In Figure \ref{Fig: Bubble Tracks} we show one such 10-frame realization for the trailer wake and empty tunnel experiments. At this trajectory concentration in a turbulent flow, meaningful time-resolved spatial derivatives are unattainable. Visual inspection of the two 10-frame realizations also gives no indication that a vortical wake is present in the trailer wake flow. Instead of relying on a spatial interpolant to visualize strong regions of rotation, we now demonstrate a complementary statistical approach using this sparse turbulent data set. 

\begin{figure}
\includegraphics[width=\textwidth]{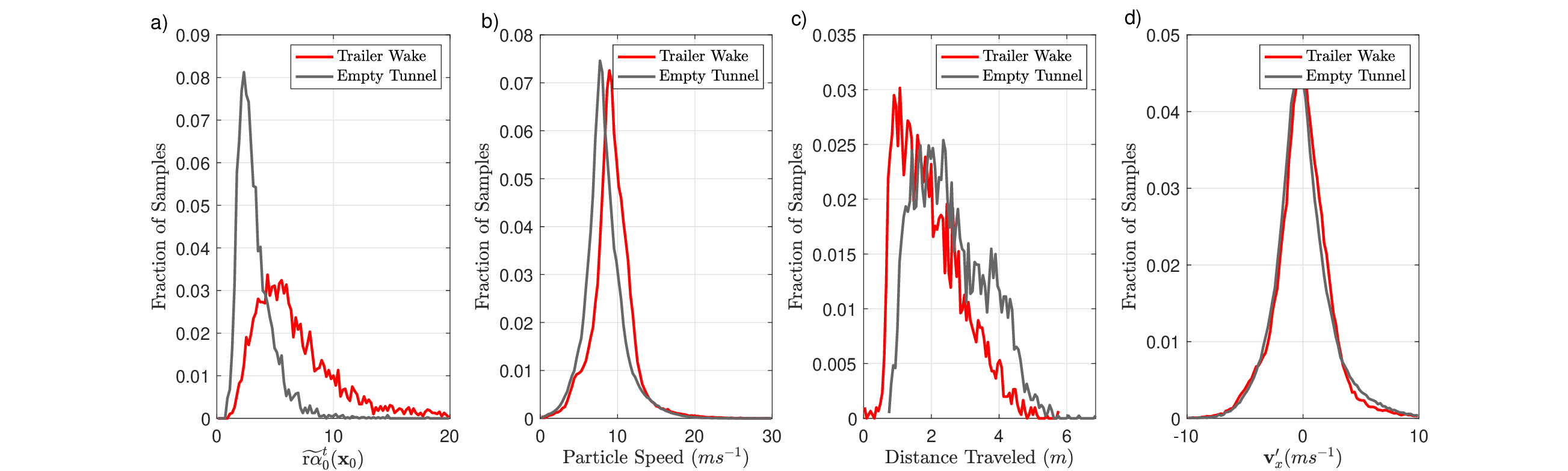}
\caption{Comparison of discrete probability histograms of frame-dependent metrics for the truck wake vortex and object-free uniform flow experiments at $\sim1\ell^{-3}$.}
\label{Fig: Truck Wake Stats}
\end{figure}

In figure \ref{Fig: Truck Wake Stats} we show the discrete probability histograms for common LPT flow diagnostics in addition to $\mathrm{r\alpha}$ analysis. The distribution of the particle speed over the entire 300 $s$ (\ref{Fig: Truck Wake Stats}b) shows a slight speed up of the flow for the trailer wake, but the bubble trajectories  travel for a shorter distance on average (\ref{Fig: Truck Wake Stats}c). The fluctuating streamwise components of bubble velocities were also nearly identical for the two flows. Accurate values for all these measurements depend on reliable, noise-free data, but given the available information from this experimental setup, none of these diagnostics present a clear picture of the wake dynamics. In fact $\mathbf{v}_x'$ cannot distinguish the two flows.

Instead, we compute 10-frame $\ralpha$ values beginning at every frame available. This gives us a distribution of $\ralpha_t^{t+10}(\boldsymbol{x}_0)$ for a wide range of $t$ and $\boldsymbol{x}_0$. From a set of trajectories at each time $t$, we then calculate the median of $\ralpha$ ($\widetilde{\ralpha}$) to quantify the general behavior of the flow in each time window. The discrete probability histogram of $\widetilde{\ralpha}$ for the two flows is shown in Figure $\ref{Fig: Truck Wake Stats}a$. The empty tunnel data is concentrated around a low peak, suggesting a relatively uniform flow domain and an accumulation of small turbulent oscillations, but no major rotational features. In contrast, the trailer wake data is distinguished by a larger range of values, with a heavy tail towards considerable rotation, and high peak $\widetilde{\ralpha}$. The physical meaning of this data suggests a more complex flow, with a higher degree of rotation, in this case because of the vortical wake. Even at this low concentration, and with this experimental methodology, we are able to identify distinguished rotating features in the flow. 

\begin{figure}
\includegraphics[width=\textwidth]{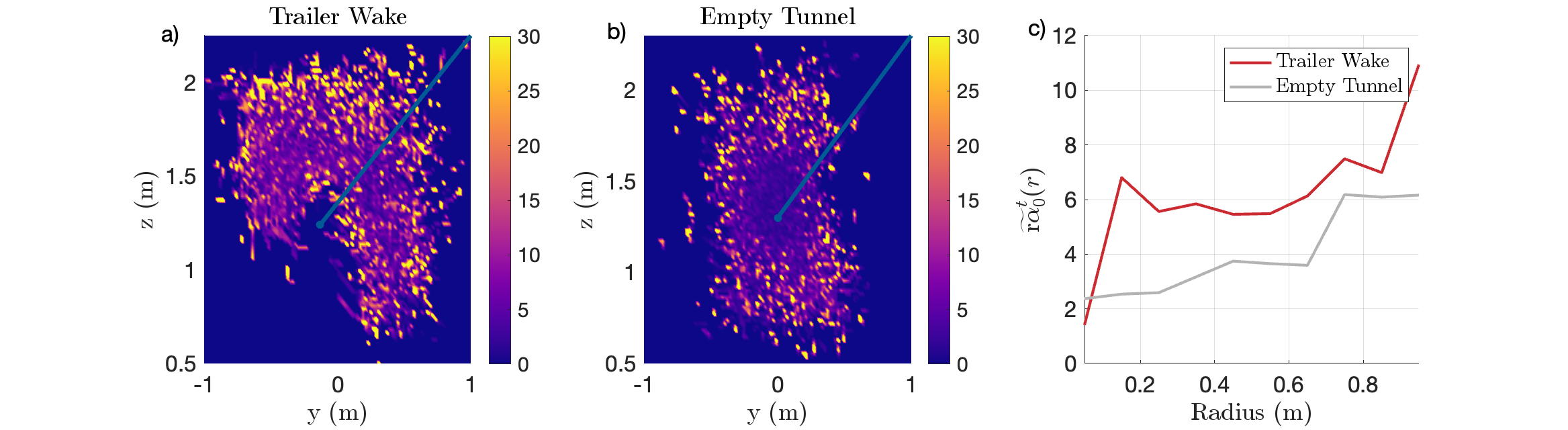}
\caption{Panels a and b show the binned median value of $\ralpha_0^t(\boldsymbol{x}_0)$ after projecting in the streamwise direction onto the y-z plane. The trailer wake has high rotation in its core, and relatively higher $\widetilde{\ralpha}_0^t$ values distributed across the flow. The empty tunnel, in contrast has a low rotational core, with higher $\widetilde{\ralpha}_0^t$ appear only on the edges of the measurement domain. This is confirmed by binning the $\ralpha_0^t$ values by radial distance from the flow center, and calculating median values ($\widetilde{\ralpha}_0^t(r)$, panel c). Note the radial centers and axes are marked by blue dots and lines in panels a and b.}
\label{Fig: Truck Wake}
\end{figure}

We can begin to distinguish the spatial features of the flow by analyzing the long-time trends in $\ralpha$. In figure \ref{Fig: Truck Wake} we project all initial particle locations to the y-z plane, and bin them in a $100\times100$ grid. We again calculate $\widetilde{\ralpha}$ for each bin and plot this in Figures \ref{Fig: Truck Wake}a-b. The trailer wake in figure \ref{Fig: Truck Wake}a has high rotation near its center, and relatively higher $\mathrm{r\alpha}$ distributed throughout the flow domain when compared with the quiescent center and lower values in \ref{Fig: Truck Wake}b. We quantify the spatial dependence of rotation on flow geometry by binning $\mathrm{r\alpha}$ in 10 bins along the radial distance from the center of the flow. We show this radial axis as a dot and line in Figure \ref{Fig: Truck Wake}a-b. Again, the median $\mathrm{r\alpha}$ values from these bins shows the trailer wake generates a much greater degree of rotation on the fluid, starting from a minimum where little data is available, with a local maximum in the vortex core, then gradually increasing as we move away in the flow. At nearly every location, the empty tunnel generates limited fluid rotation with lower $\mathrm{r\alpha}$ values.

The results here provide a good baseline for experimental LPT data that was not designed with $\ralpha$ analysis in mind. As future experiments may be designed that consider $\mathrm{r\alpha}$ requirements, further improvement on the identification and analysis of coherent rotational flow features is expected. Furthermore, given the connections between $\ralpha$, curvature (e.g, Section \ref{Sec: Rotation}), and rotation times, concerted future effort may help connect classical notions of geometry, eddy turnover times, spectral analysis and sparse trajectory data in an objective manner, in addition to designating strongly and weakly rotating regions of a flow.

\section{Conclusion and Outlook}

Objective coherent structure identification tools are important in a wide range of scientific and industrial fields. Methods that work equally well for gridded and sparse Lagrangian data allow meaningful comparisons between model and observations, and can provide new insights for experimental studies. In this work, we have mathematically derived new frame-independent diagnostics of trajectory rotation and stretching well-suited for coherent structure identification in sparse experimental trajectory data. Through multiple systematic comparisons with existing Lagrangian flow diagnostics, we have shown that:

\begin{itemize}
\item $r\mathrm{\overline{TSE}}$ and $\mathrm{r\alpha}$ are objective metrics that can identify elliptic and hyperbolic Lagrangian coherent structures in highly unsteady flows, and that work equally well in flows with large or small bulk translation and rotation;
\item $r\mathrm{\overline{TSE}}$ and $\mathrm{r\alpha}$ work well in many of the situations where the quasi-objective diagnostics $\mathrm{\overline{TSE}}$ and $\mathrm{\overline{TRA}}$ have previously been successful, including ocean drifter applications;
\item $\mathrm{r\alpha}$ can designate regions with strong rotation in 2D and 3D flows for sparse data beyond the capabilities of existing techniques. 
\item $\ralpha$ values maintain accuracy and and interpretability at extremely sparse sampling, especially when compared with common frame-dependent diagnostics; and
\item $r\mathrm{\overline{TSE}}$ and $\mathrm{r\alpha}$ provide an avenue for meaningful LPT analysis and post-processing that does not require assumptions of steady state flows or data interpolation to an Eulerian grid.
\end{itemize}

To assist future practitioners, we have also developed two guidelines based on the results of the present research and our experiences while conducting it:

\begin{itemize}
\item If the trajectory concentration is below one $\ell^{-2}$ or $\ell^{-3}$, we recommend using time-averaged values of the experiment-derived $\overline{\mathbf{v}}$, $\overline{\omega}$ and $\overline{\mathbf{x}}$ 
\item As the trajectory concentration decreases, the continuity of individual trajectory data becomes more important as each data point contributes more to spatial averages. It is then beneficial to select observation windows or trajectory sets for which trajectories are continuously present to prevent un-physical fluctuations in $\mathbf{v}_d$ caused by intermittent trajectory data.
\end{itemize}

Being able to identify coherent flow features using $r\mathrm{\overline{TSE}}$ and $\mathrm{r\alpha}$ for the analysis of unsteady flows in a physically meaningful, frame-indifferent manner will help advance our study of fluid dynamics in large and natural domains. Further applications of these methods, such as the similar-time-series analytic techniques used by \citet{Aksamit2023a}, is expected to open the door for an even wider range of experiments over a vast range of scales that can harness the Lagrangian nature of turbulent structures. 

\backsection[Acknowledgements]{The authors would like to express their gratitude to three anonymous reviewers whose comments helped improve the manuscript.}

\backsection[Funding]{This research received no specific grant from any funding agency, commercial or not-for-profit sectors}

\backsection[Declaration of interests]{{\bf Declaration of Interests}. The authors report no conflict of interest.}

\backsection[Data availability statement]{The LPT wind tunnel data that support the findings of this study are available from the authors directly. The turbulent DNS data can be found at \href{https://doi.org/10.15454/GLNRHK}{https://doi.org/10.15454/GLNRHK}. AVISO ocean surface current data can be found at \href{https://doi.org/10.48670/moi-00145}{https://doi.org/10.48670/moi-00145}. Ocean drifter data can be accessed through \href{https://www.aoml.noaa.gov/phod/gdp/index.php}{https://www.aoml.noaa.gov/phod/gdp/index.php}.}

\backsection[Author ORCIDs]{N. Aksamit, \href{https://orcid.org/0000-0002-2610-7258}{https://orcid.org/0000-0002-2610-7258}; A. Encinas-Bartos, \href{https://orcid.org/0000-0002-4203-9128}{https://orcid.org/0000-0002-4203-9128}; G. Haller, \href{https://orcid.org/0000-0003-1260-877X}{https://orcid.org/0000-0003-1260-877X}; D. Rival, \href{https://orcid.org/0000-0001-7561-6211}{https://orcid.org/0000-0001-7561-6211}.}

\appendix \section{Objectivity of $\boldsymbol{v}_d$}\label{app:1}

We are interested in how $\boldsymbol{v}_d$ changes under Euclidean transformations of the form 
\begin{align}
\boldsymbol{x}(t)=Q(t)\boldsymbol{y}(t)+\boldsymbol{b}(t).
\end{align} Under this transformation $\boldsymbol{y}(t)=Q^T(t)(\boldsymbol{x}(t)-\boldsymbol{b}(t))$. Differentiating with respect to time gives us the transformation of the velocity

\begin{align}
\hat{\boldsymbol{v}}(t)=Q^T(t)[\boldsymbol{v}(t)-\dot{\boldsymbol{b}}(t)-\dot{Q}(t)Q^T(t)(\boldsymbol{x}(t)-\boldsymbol{b}(t))]
\end{align}
and
\begin{align}
\hat{\boldsymbol{v}}(t)-\hat{\overline{\boldsymbol{v}}}(t)=Q^T(t)[\boldsymbol{v}(t)-\overline{\boldsymbol{v}}(t)-\dot{Q}(t)Q^T(t)(\boldsymbol{x}(t)-\overline{\boldsymbol{x}}(t))].
\end{align}

In the new reference frame, $\Theta$ transforms as

\begin{align}
\begin{split}
\hat{\Theta}&=\overline{|Q^T(\boldsymbol{x}-\overline{\boldsymbol{x}})|^2I-Q^T(\boldsymbol{x}-\overline{\boldsymbol{x}})\otimes Q^T(\boldsymbol{x}-\overline{\boldsymbol{x}})}\\
&=\overline{|(\boldsymbol{x}-\overline{\boldsymbol{x}})|^2I-Q^T(\boldsymbol{x}-\overline{\boldsymbol{x}})(\boldsymbol{x}-\overline{\boldsymbol{x}})^TQ}\\
&=\overline{Q^T|(\boldsymbol{x}-\overline{\boldsymbol{x}})|^2IQ-Q^T(\boldsymbol{x}-\overline{\boldsymbol{x}})(\boldsymbol{x}-\overline{\boldsymbol{x}})^TQ}\\
&=Q^T[\overline{|(\boldsymbol{x}-\overline{\boldsymbol{x}})|^2I-(\boldsymbol{x}-\overline{\boldsymbol{x}})\otimes(\boldsymbol{x}-\overline{\boldsymbol{x}})}]Q\\
&=Q^T\Theta Q.
\end{split}
\end{align}

We also have the useful identities
\begin{equation}
\dot{Q}Q^T(\boldsymbol{x}-\overline{\boldsymbol{x}})=\dot{\boldsymbol{q}}\times(\boldsymbol{x}-\overline{\boldsymbol{x}})
\end{equation}
and
\begin{align}
\begin{split}
\Theta\dot{\boldsymbol{q}}&=\overline{|\boldsymbol{x}-\overline{\boldsymbol{x}}|^2I-(\boldsymbol{x}-\overline{\boldsymbol{x}})\otimes(\boldsymbol{x}-\overline{\boldsymbol{x}})}\dot{\boldsymbol{q}}\\
&=\overline{|\boldsymbol{x}-\overline{\boldsymbol{x}}|^2\dot{\boldsymbol{q}}-(\boldsymbol{x}-\overline{\boldsymbol{x}})(\boldsymbol{x}-\overline{\boldsymbol{x}})^T\dot{\boldsymbol{q}}}\\
&=\overline{|\boldsymbol{x}-\overline{\boldsymbol{x}}|^2\dot{\boldsymbol{q}}-(\boldsymbol{x}-\overline{\boldsymbol{x}})\left((\boldsymbol{x}-\overline{\boldsymbol{x}})\cdot\dot{\boldsymbol{q}}\right)}.\\
\end{split}
\end{align}

Under the frame change, we find $\hat{\overline{\boldsymbol{\omega}}}=\hat{\Theta}^{-1}\overline{(\boldsymbol{y}-\overline{\boldsymbol{y}})\times(\hat{\boldsymbol{v}}-\hat{\overline{\boldsymbol{v}}})}.$ So that

\begin{align}
\begin{split}
\hat{\Theta}\hat{\overline{\boldsymbol{\omega}}} &= \overline{(\boldsymbol{y}-\overline{\boldsymbol{y}})\times(\hat{\boldsymbol{v}}-\hat{\overline{\boldsymbol{v}}})} \\
&= \overline{Q^T(\boldsymbol{x}-\overline{\boldsymbol{x}})\times Q^T(\boldsymbol{v}-\overline{\boldsymbol{v}}-\dot QQ^T(\boldsymbol{x}-\overline{\boldsymbol{x}}))} \\
&=Q^T\overline{(\boldsymbol{x}-\overline{\boldsymbol{x}})\times (\boldsymbol{v}-\overline{\boldsymbol{v}}) -(\boldsymbol{x}-\overline{\boldsymbol{x}})\times \dot{\boldsymbol{q}}\times(\boldsymbol{x}-\overline{\boldsymbol{x}})} \qquad \text{by (A5)}\\
&=Q^T\overline{(\boldsymbol{x}-\overline{\boldsymbol{x}})\times (\boldsymbol{v}-\overline{\boldsymbol{v}})} -Q^T\overline{\dot{\boldsymbol{q}}|\boldsymbol{x}-\overline{\boldsymbol{x}}|^2 + (\boldsymbol{x}-\overline{\boldsymbol{x}})(\dot{\boldsymbol{q}}\cdot(\boldsymbol{x}-\overline{\boldsymbol{x}}))} \\
&=Q^T\overline{(\boldsymbol{x}-\overline{\boldsymbol{x}})\times (\boldsymbol{v}-\overline{\boldsymbol{v}})} -Q^T\Theta\dot{\boldsymbol{q}} \qquad \text{from (A6)} \\
&=Q^T[\Theta\overline{\boldsymbol{\omega}} -\Theta\dot{\boldsymbol{q}}].
\end{split}
\end{align}

Since $\hat{\Theta}=Q^T\Theta Q$, we have $Q^T\Theta Q\hat{\overline{\boldsymbol{\omega}}} = Q^T[\Theta\overline{\boldsymbol{\omega}} - \Theta\dot{\boldsymbol{q}}]$ and
\begin{align}
\hat{\overline{\boldsymbol{\omega}}}=Q^T[\overline{\boldsymbol{\omega}}-\dot{\boldsymbol{q}}].
\end{align}

These identities together lead to
\begin{align}
\begin{split}
\hat{\boldsymbol{v}}_d(\boldsymbol{y}(t))&=Q^T[\boldsymbol{v}-\overline{\boldsymbol{v}}-\dot{Q}Q^T(\boldsymbol{x}-\overline{\boldsymbol{x}})]-\hat{\overline{\boldsymbol{\omega}}}\times(Q^T(\boldsymbol{x}-\overline{\boldsymbol{x}}))\\
&=Q^T[\boldsymbol{v}-\overline{\boldsymbol{v}}-\dot{Q}Q^T(\boldsymbol{x}-\overline{\boldsymbol{x}})]-Q^T[\overline{\boldsymbol{\omega}}-\dot{\boldsymbol{q}}]\times(Q^T(\boldsymbol{x}-\overline{\boldsymbol{x}})) \qquad \text{by (A8)}\\	
&=Q^T[\boldsymbol{v}-\overline{\boldsymbol{v}}-\dot{Q}Q^T(\boldsymbol{x}-\overline{\boldsymbol{x}})]-Q^T[\overline{\boldsymbol{\omega}}\times(\boldsymbol{x}-\overline{\boldsymbol{x}})] + Q^T\dot{\boldsymbol{q}}\times(\boldsymbol{x}-\overline{\boldsymbol{x}})\\
&=Q^T[\boldsymbol{v}-\overline{\boldsymbol{v}}-\dot{Q}Q^T(\boldsymbol{x}-\overline{\boldsymbol{x}})-\overline{\boldsymbol{\omega}}\times(\boldsymbol{x}-\overline{\boldsymbol{x}}) + \dot{Q}Q^T(\boldsymbol{x}-\overline{\boldsymbol{x}})]\\
&=Q^T[\boldsymbol{v}-\overline{\boldsymbol{v}}-\overline{\boldsymbol{\omega}}\times(\boldsymbol{x}-\overline{\boldsymbol{x}})]\\
&=Q^T\boldsymbol{v}_d(\boldsymbol{x}(t)).
\end{split}
\end{align}

\section{Sensitivity of reconstructed r$\alpha$-fields to choice of interpolant}
\label{Appendix: Interpolant}

\begin{figure}
 \includegraphics[width=\textwidth]{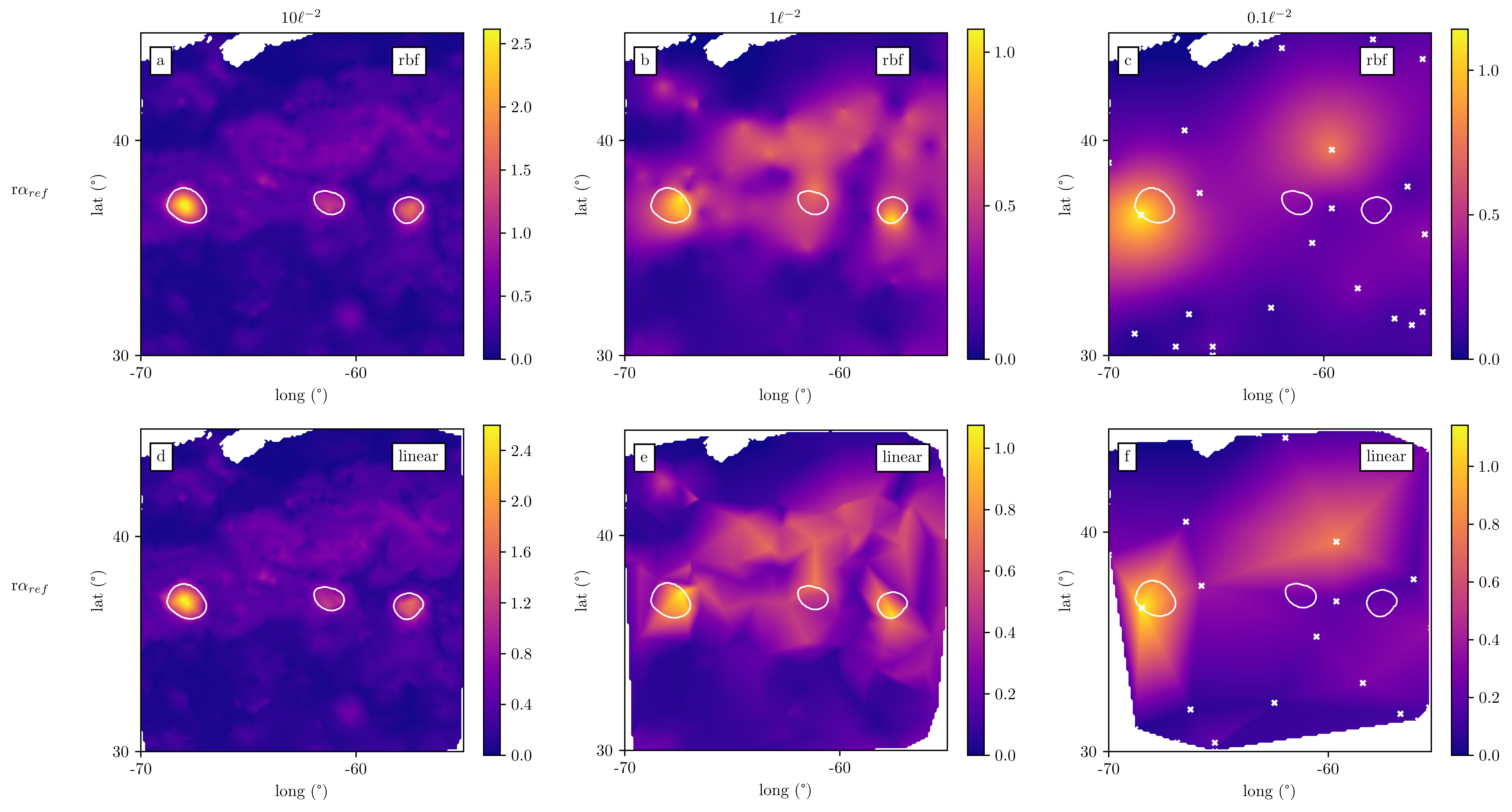}
 \caption{Comparison of $\ralpha_{ref}$ field for different interpolation schemes. The upper row (panels a-c) shows the $\ralpha_{ref}$ field reconstructed using radial basis functions (rbf) for different trajectory densities. The lower row (panels d-f) shows the $\ralpha_{ref}$ field reconstructed using standard linear interpolation for different trajectory densities. The white contours indicate the vortex boundary extracted from LAVD-field at full resolution.}
 \label{fig:Interpolant Sensitivity}
 \end{figure}

In order to reconstruct the $\ralpha$-field, we need to interpolate the sparse $\ralpha$ values over a regular meshgrid. The exact values of the reconstructed $\ralpha$-field depend on the specific interpolation method. In this appendix, we perform a sensitivity analysis of the interpolated $\ralpha$-field as we progressively subsample the trajectory density in the AVISO dataset (Section \ref{sec:AVISO}. Fig. \ref{fig:Interpolant Sensitivity} compares the reconstructed $\ralpha_{ref}$ field for different trajectory densities. For a trajectory density of $10\ell^{-2}$ the reconstructed $\ralpha_{\mathrm{ref}}$ field remains nearly unaffected by the interpolation scheme. For low trajectory densities (1 - $0.1\ell^{-2}$), the exact boundaries of features in the $\ralpha_{\mathrm{ref}}$ field vary with on the choice of interpolation scheme, but we expect the overall topology of the $\ralpha_{\mathrm{ref}}$ field to be robust with respect to the employed interpolation method as they are the result of raw values, and are not generated by the interpolants. Specifically, local maxima of the $\ralpha_{\mathrm{ref}}$ clearly persist irrespective of the chosen interpolation method. Similarly to previous studies by \cite{Encinas-Bartos2022}, we recommend using rbf interpolation since it favors elliptic and smooth structures.

\bibliographystyle{jfm}
\bibliography{references}
\end{document}